\documentclass[fleqn,usenatbib]{mnras}

\usepackage{newtxtext,newtxmath}

\usepackage[T1]{fontenc}
\usepackage{ae,aecompl}

\usepackage{amsmath}
\usepackage{natbib}
\usepackage{lscape}
\usepackage{color}
\usepackage{subfigure}
\usepackage{amssymb}
\usepackage{amsmath}
\usepackage{url}
\usepackage{amsfonts}
\usepackage{amsbsy}
\usepackage{graphicx}
\usepackage{subfigure}
\usepackage{verbatim}
\usepackage{multicol}
\usepackage[normalem]{ulem}

\usepackage{graphicx}	
\usepackage{amsmath}	
\usepackage{amssymb}	

\title[The OH Megamaser Galaxy IRAS17526+3253]{Gemini IFU, VLA, and HST observation of the OH Megamaser Galaxy IRAS17526+3253\footnote{Based partly on observations made with the NASA/ESA Hubble Space Telescope, obtained at the Space Telescope Science Institute, which is operated by the Association of Universities for Research in Astronomy, Inc., under NASA contract NAS 5-26555.}}

\author[Sales et al.]{
Sales, Dinalva A.$^{1}$\thanks{E-mail: dinalvaires@gmail.com}
Robinson, A.$^{2,3}$; 
Riffel, R. A.$^{4}$; 
Storchi-Bergmann, T.$^{5}$; \\
\newauthor Gallimore, J. F.$^{6}$;
Kharb, P.$^{7}$;
Baum, S.$^{8,3}$;
O'Dea, C.$^{8,2}$; 
Hekatelyne, C. $^{4}$;
Ferrari, F.$^{1}$\\
$^{1}$Universidade Federal do Rio Grande, Instituto de Matem\'{a}tica, Estat\'{i}stica e F\'{i}sica, 8 km It\'{a}lia Av., Rio Grande, RS, Brazil\\
$^2$School of Physics and Astronomy, Rochester Institute of Technology, 84 Lomb Memorial Drive, Rochester, NY 14623, USA\\
$^3$Chester F. Carlson Center for Imaging Science, Rochester Institute of Technology, 54 Lomb Memorial Drive, Rochester, NY 14623, USA\\
$^{4}$Universidade Federal de Santa Maria, Departamento de F\'{i}sica, Centro de Ci\^{e}ncias Naturais e Exatas, 97105-900 Santa Maria, RS, Brazil\\
$^5$Departamento de Astronomia, Universidade Federal do Rio Grande do Sul. 9500 Bento Gon\c calves, Porto Alegre, 91501-970, Brazil\\
$^6$Department of Physics and Astronomy, Bucknell University, Lewisburg, PA 17837, USA\\
$^7$National Centre for Radio Astrophysics, Tata Institute of Fundamental Research, Pune University Campus, Ganeshkhind, Pune, India\\
$^8$Department of Physics and Astronomy, University of Manitoba, Winnipeg, MB, R3T 2N2, Canada}

\date{Accepted XXX. Received YYY; in original form ZZZ}

\pubyear{2017}

\begin{document}
\label{firstpage}
\pagerange{\pageref{firstpage}--\pageref{lastpage}}
\maketitle

\begin{abstract}

We present a multiwavelength study of the OH megamaser galaxy (OHMG)
IRAS17526+3253, based on new Gemini Multi-Object Spectrograph Integral Field Unit (GMOS/IFU) observations,
Hubble Space Telescope F814W and H$\alpha$+[N{\sc ii}] images, and archival 2MASS and 1.49GHz VLA data. 
The HST images clearly reveal a mid-to-advanced stage major merger
whose northwestern and southeastern nuclei have a projected 
separation of $\sim$8.5kpc. Our HST/H$\alpha$+[N{\sc ii}] image shows regions of ongoing star-formation 
across the envelope on $\sim$10kpc scales, which are aligned with radio features, supporting the interpretation that the radio  
emission originates from star-forming regions. The measured H$\alpha$ luminosities 
imply that the unobscured star-formation rate is  $\sim$10-30\,M$_{\odot}$yr$^{-1}$. 
The GMOS/IFU data reveal two structures in northwestern separated by 850\,pc and by  
a discontinuity in the velocity field of $\sim$~200~km~s$^{-1}$. 
We associate  the blue-shifted and red-shifted components with, respectively, the distorted disk of northwestern and
tidal debris, possibly a tail originating in southeastern.
Star-formation is the main ionization source in both components, which have SFRs of 
$\sim$2.6-7.9\,M$_{\odot}$yr$^{-1}$ and $\sim$1.5-4.5\,M$_{\odot}$yr$^{-1}$, respectively. Fainter line 
emission bordering these main components is consistent with shock ionization at a velocity $\sim$200~km~s$^{-1}$ 
and may be the result of an interaction between the tidal tail and the northwestern galaxy's disk. 
IRAS17526+3253 is one of only a few systems known to host both luminous OH and H$_{2}$O masers. 
The velocities of the OH and H$_{2}$O maser lines suggest that they are associated 
with the northwestern and southeastern galaxies, respectively.

\end{abstract}




\section{Introduction}\label{sec:introd}

Galaxy mergers are generally believed to play an important role in galaxy evolution, with 
the assembly of massive haloes via multiple mergers being the central mechanism for structure 
formation in $\Lambda$CDM  cosmologies \citep[e.g.,][]{Blumenthal1984, Springel2005, Boylan-Kolchin2009, Angulo2012}. 
Observational evidence and cosmology simulations suggest that the most
massive galaxies are formed in major mergers, whilst the growth of intermediate mass 
galaxies is largely driven by minor mergers or cold gas accretion 
\citep[e.g., reviews by][]{Silk2012, Kormendy2013, Conselice2014}.
In this context, the population of optically faint but infrared luminous galaxies discovered 
by the Infrared Astronomical Satellite (IRAS) in the 1980's \citep{Soifer1984} is of great 
interest, since it includes many examples of gas-rich major mergers, or strong interactions 
\citep[e.g.,][]{Sanders1988, Veilleux2002, Haan2011}. The immense infrared luminosities of these
objects arise from dust heated by embedded starbursts and/or active galactic nuclei (AGN), 
both of which are thought to be triggered by gas inflows driven by tidal torques generated 
by the interaction \citep[e.g.,][]{Barnes1992, Mihos1996, Hopkins2006}. 
Two main subdivisions of this population are recognized:  Luminous infrared galaxies (LIRGs) 
have far infrared luminosities of $L_{FIR} \gtrsim 10^{11} L_{\odot}$, whereas the most luminous 
systems, ultra-luminous infrared galaxies (ULIRGs), have quasar-like luminosities 
of $L_{FIR} \gtrsim 10^{12} L_{\odot}$ \citep[see][for a review]{Sanders1996}. It has been 
proposed that (U)LIRGs represent different phases in the evolution of gas rich mergers, 
with LIRGs evolving into ULIRGs as the merger progresses, eventually followed  by the 
emergence of a luminous AGN in a massive elliptical host, as the remaining circum-nuclear 
gas and dust is dispersed by starburst and AGN-induced outflows \citep{Sanders1988, Veilleux2002}.

Many studies of local (U)LIRGs broadly support this idea. For example, merger morphology  
correlates with FIR luminosity, with advanced mergers becoming more prevalent at higher 
luminosities \citep[e.g.,][]{Sanders1988, Veilleux2002, Hwang2010, Haan2011, Carpineti2015}, 
and the fraction hosting powerful active galactic nuclei (AGN) increases with both FIR 
luminosity, and merger stage \citep[e.g.,][]{Veilleux2002, Veilleux2009, Nardini2010}.  
At higher redshifts ($z\sim 1 - 2$) (U)LIRGs are much more common than in the local 
universe  \citep[e.g.,][]{Caputi2007}, 
and similar morphological trends are seen, although with a wider range of merger states 
at high luminosity \citep{Kartaltepe2012}. Beyond $z > 2$, the morphological properties of 
the most luminous submillimeter galaxies (SMGs) are also consistent with mergers and interacting 
systems \citep[e.g.,][]{Tacconi2008, Engel2010, Hayward2013}.
Massive outflows in neutral, ionized and molecular gas have been detected in an most 
ULIRGs, both in the local universe \citep[e.g.,][]{Heckman2000, Rupke2002, Martin2005, Feruglio2010, Sturm2011, Rupke2013, Spoon2013, Veilleux2013, Cicone2014} 
and at higher redshifts in both ULIRGs \citep[e.g.,][]{George2014} and SMGs \citep[e.g.,][]{Banerji2011}.
(U)LIRGs are therefore crucial to understanding the role of  mergers in galaxy evolution 
and black hole growth. 

Integral field spectroscopy is a powerful tool with which to study 
these complex systems, providing spatially resolved information on gas and stellar kinematics 
and a range of spectral diagnostics. Using such information (for example) gas flows can be 
mapped and contributions of various ionization mechanisms (AGN or stellar photoionization, 
or shocks) can be inferred \citep[e.g.,][]{Monreal2006,  Monreal2010, Rich2012, Rich2015}.

Here, we present a high-spatial resolution analysis of the morphology, gaseous excitation and kinematics 
of the brightest radio nucleus of the merger system IRAS17526+3253. This system, which has 
a redshift $z=0.025$ \citep{RC3}, was first identified as a possible interacting galaxy by 
\citet{Andreasian1994}. In the Digitized Sky Survey image, it appears as a warped, elongated 
structure oriented approximately southeastern -- northwestern, and has two bright knots separated 
by $\approx 20\arcsec$. 

It has a far-infrared (FIR) luminosity, as calculated from the 60 and 100$\mu$m IRAS fluxes, of $L_{FIR} = 10^{11.19} $\,L$_{\odot}$ 
(taken from The Imperial IRAS-FSC Redshift Catalogue; \citet[IIFSCz][]{Wang2009}), which places it just above
the lower limit for LIRGs. 
An optical spectrum obtained using a 2\arcsec\ wide slit by \citet{Baan1998} shows line 
ratios characteristic of a starburst spectrum. 

\citet{Baan2006} observed IRAS17526+3253 with the VLA A-array in the L and C bands (1.49 
and 4.9 GHz), finding that the radio source consists of a linear chain of knots, with 3 
main components, spanning $\sim 30$\arcsec\ ($\sim 15$\,kpc) and extending along PA 119$^{\circ}$, 
closely aligned with the optical PA.  
These authors used three indicators to distinguish between AGN and starburst activity as 
the origin of the radio emission: the brightness temperature at 4.85 GHz, the FIR-radio flux 
ratio at 4.85 GHz, and the 1.4--4.8\,GHz spectral index, $\alpha^{1.4GHz}_{4.8GHz}$. Based on these, all three main components 
were classified as starburst-powered, although in the case of the northwestern component
the peak spectral index has a value $S \propto \nu^{-0.25\pm\,0.22}$ \citep{Baan2006}, 
which may indicate a typical of AGN or the flat spectrum is due to te free-free emission \citep{Gioia1982,Condon1983,Condon1991}. 







\citet{Garwood1987} were the first to look for the 1665\,MHz and 1667\,MHz OH maser lines 
in IRAS17526+3253, but reported a non-detection. A detection was subsequently reported 
by \citet{Martin1989a}, with an integrated luminosity 
log($L_{\rm OH}/{\rm L_\odot}$)=0.99 and recession velocity $\approx 7500$km\,s$^{-1}$ \citep[see also][]{Martin1989b}, although
\citet{McBride2013} were unable to confirm the detection in more recent Arecibo Telescope observations due to strong radio frequency interference. 
\citet{Martin1989b} also observed the H\,I 21\, cm emission line in IRAS17526+3253, at a velocity $\approx 7800$km\,s$^{-1}$, close to the
heliocentric systemic velocity ($7798\pm  9$\,km\,s$^{-1}$; \citealp{RC3}) but redshifted by $\approx 300$\,km\,s$^{-1}$ with respect to the OH maser. 
Molecular line observations by \citep{Baan2008} detected CO emission in  the $^{12}$CO(1-0) and $^{12}$CO(2-1) transitions, revealing (in the latter) two velocity components, a broad  peak at $\approx 7500$km\,s$^{-1}$ and a narrower peak at $\approx 7800$km\,s$^{-1}$. IRAS17526+3253 is also known to host a 22\,GHz H$_{2}$O kilo-maser source \citep{Wagner2013}, 
which exhibits a cluster of three narrow ($\approx 10$km\,s$^{-1}$) features at $\approx 7800$km\,s$^{-1}$, close to the systemic velocity. 
The H$_{2}$O maser lines have an integrated luminosity $\sim 360$\,L$_\odot$ and possibly arise from a region of 
shocked gas. 


Our study is based on new HST/ACS images and integral field spectroscopy obtained with the 
GMOS/IFU at the Gemini North Telescope. The  paper is organized as follows. In Section \ref{sec:data} 
we describe the data reduction procedures and the measurements derived from the HST imaging 
and GMOS/IFU data. Section \ref{sec:results} describes our results on the overall morphology 
of the IRAS17526+3253 system, and the gas excitation and kinematics within a $5\farcs1 \times  3\farcs4$ 
field of view covering the northwestern of the two optical nuclei. We discuss the implications 
of our results in Section \ref{sec:discussion}. The main results and conclusions are summarized 
in Section \ref{sec:summary}. Throughout this paper, we adopt the Hubble constant as 
H$_0$ = 70.5 km s$^{-1}$ Mpc$^{-1}$, $\Omega_{\Lambda}$=0.73, and $\Omega_m$=0.27 
\citep{Ade2013,Lahav2014}, corresponding to a scale at the galaxy of 497 pc arcsec$^{-1}$.


\section{Observation and Data Reduction}\label{sec:data}
\subsection{Hubble Space Telescope Images}\label{sec:halpha_obser}


We acquired HST images of IRAS17526+3253 with the Advanced Camera for Surveys (ACS), as 
part of a snapshot program of a large sample of OHMGs (Program ID 11604; PI: D.J. Axon). 
The ACS wide-field channel (WFC) with broad (F814W), narrow (FR656N) and medium (FR914M) 
band filters was used.

The F814W filter was required to map the continuum morphology of the host galaxy. The 
ramp filter images were selected to cover the H$\alpha$ line in the narrow-band filter and the 
nearby continuum in the medium-band filter. The FR656N ramp filter 
includes the H$\alpha$ and the [N{\sc ii}]$\lambda 6548, 83$ lines at the redshift of IRAS17526+3253. The total 
integration times were 600s in the broad-band (I) F814W filter, 200s in the medium-band 
FR914M filter and 600s in the narrow band H$\alpha$ FR656N filter.

We used the pipeline image products for further processing with the IRAF\footnote{IRAF is  
distributed by the National Optical Astronomy Observatory, which is operated by the Association  
of Universities for Research in Astronomy (AURA), Inc., under cooperative agreement with  
the National Science Foundation.} package. We removed cosmic rays from individual 
images using the IRAF task $lacos_{im}$ \citep{Vandokkum2001}. Standard IRAF tasks 
were used to yield the final reduced images (see figure~\ref{fig:morphology}).

The continuum-free H$\alpha+$[N{\sc ii}] image of IRAS17526+3253 was derived using the following
procedure \citep[see][]{Hoopes1999,Rossa2000,Rossa2003,Sales2015}. First, we calculated the mean 
ratio of count rates, FR656N$/$FR914M, for several foreground stars that appear in both images. 
We then scaled the FR914M image by the mean count rate ratio and subtracted it from the FR656N image. 
The resulting, continuum-subtracted H$\alpha+$[N{\sc ii}] image was checked to verify that the 
residuals at the positions of the foreground stars were negligible compared to the expected noise level.
This method gives typical uncertainties of 5-10\% \citep[see][]{Hoopes1999,Rossa2000,Rossa2003}.

\subsection{VLA Radio Data}\label{sec:radio_obser}

\begin{figure*}
\centering
\includegraphics[scale=0.6]{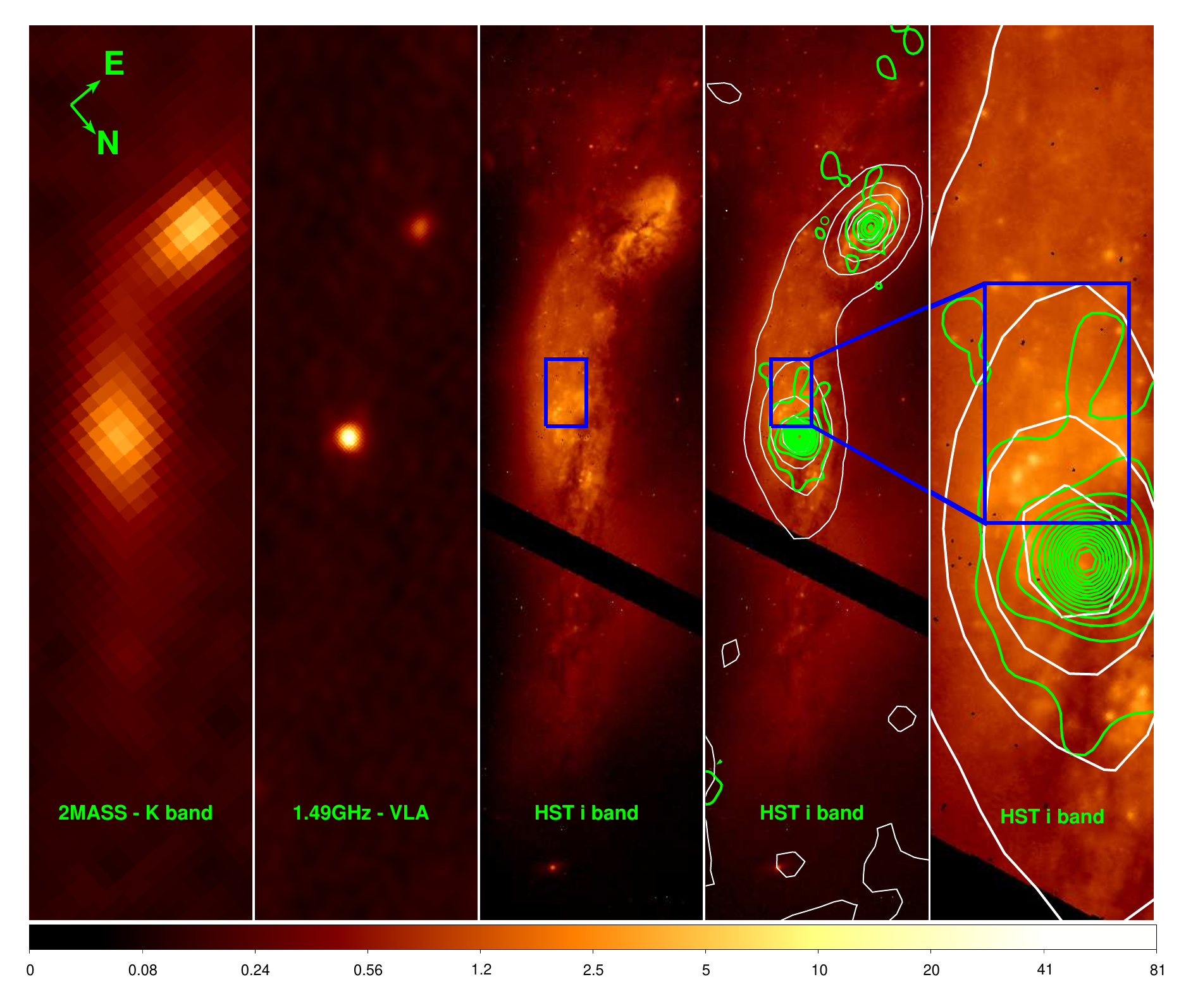}
\caption{Comparison of near infra-red, radio and optical morphology. From left to right: 2MASS K band image; 1.49 GHz VLA image; HST/ACS F814W image, with IFU field shown in blue;  HST/ACS F814W image with 1.49 GHz radio and  K band contours overplotted in green and white, respectively; zoomed in view of the northwestern nucleus in the HST i-band image. Scale bar is in arbitrary units.}
\label{fig:2mass-radio}
\end{figure*}

IRAS17526+3253 was observed at 1.49\,GHz  using the VLA A-array configuration
on March 7, 1990 (Project code: AM293). These data have previously been presented by \citet{Baan2006}. 
We re-reduced these data following standard procedures in the Astronomical Imaging Processing System (AIPS). 
The final 1.49 GHz VLA image with a beam-size of 1.35 arcsec x 1.11 arcsec at a PA = -39 degrees 
(figure \ref{fig:2mass-radio}) was obtained after several iterations of phase and amplitude self-calibration, 
using the AIPS tasks CALIB and IMAGR.

The final resultant rms noise in the image was  67~$\mu$Jy~beam$^{-1}$. 
The radio image showing two unresolved cores, has been used to derive the astrometry of the  HST, 
Gemini/IFU, and 2MASS Ks images

\subsection{Gemini Multi-Object Spectrograph Integral Field Unit Data}\label{sec:ifu_obser}

\begin{figure*}
\centering
\begin{tabular}{cc}
\includegraphics[scale=0.6]{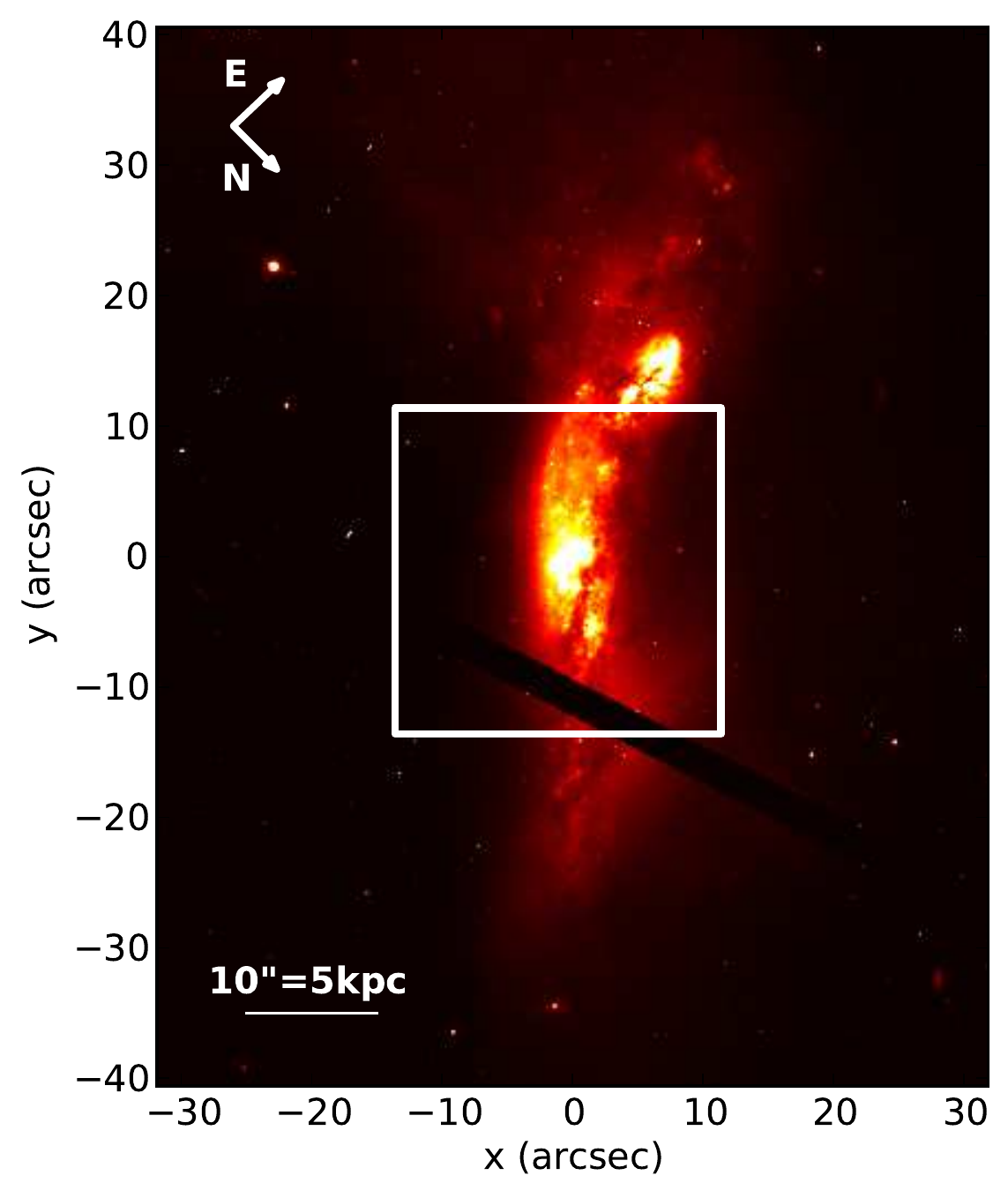}&
\includegraphics[scale=0.6]{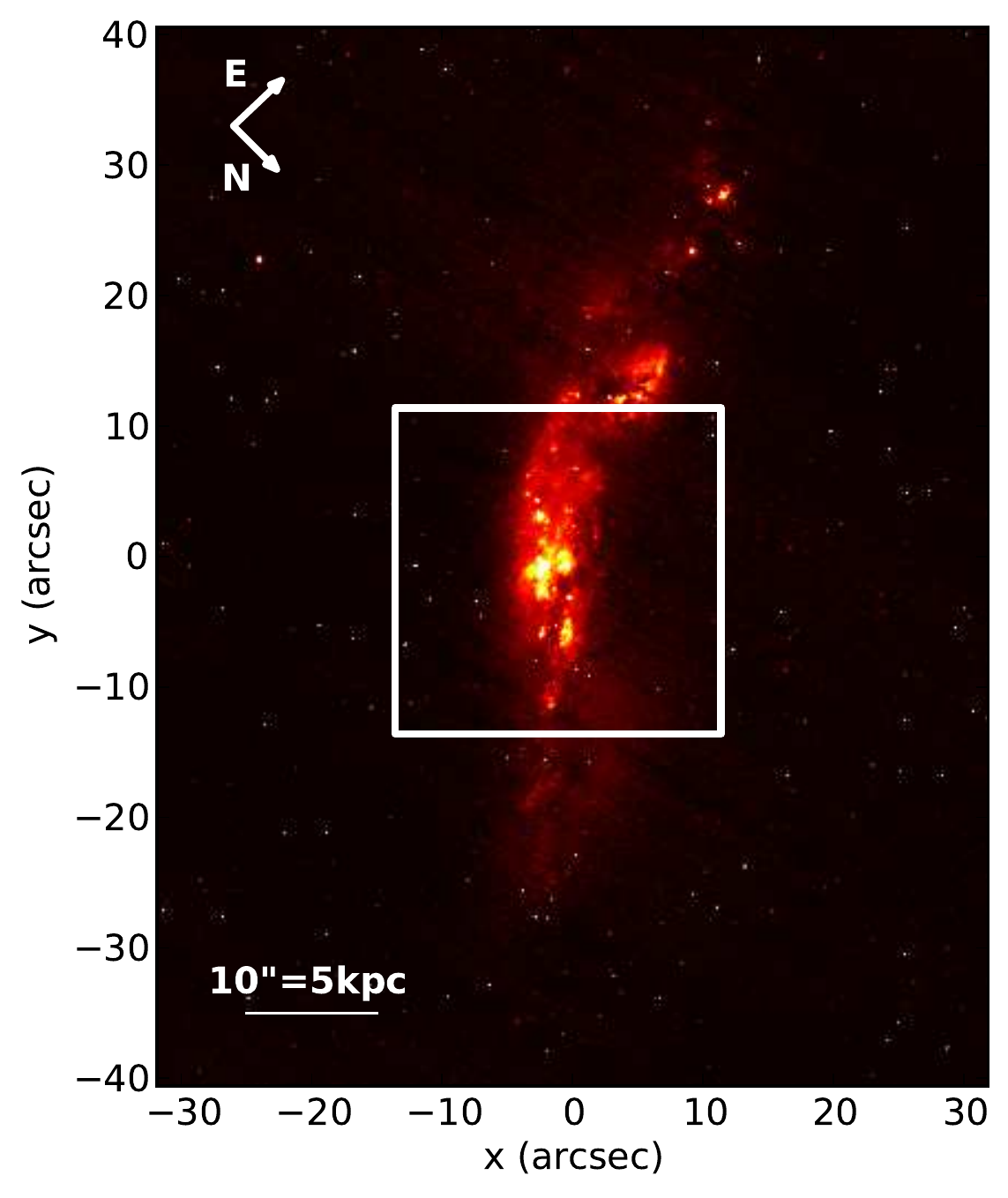}\\
\includegraphics[scale=0.56]{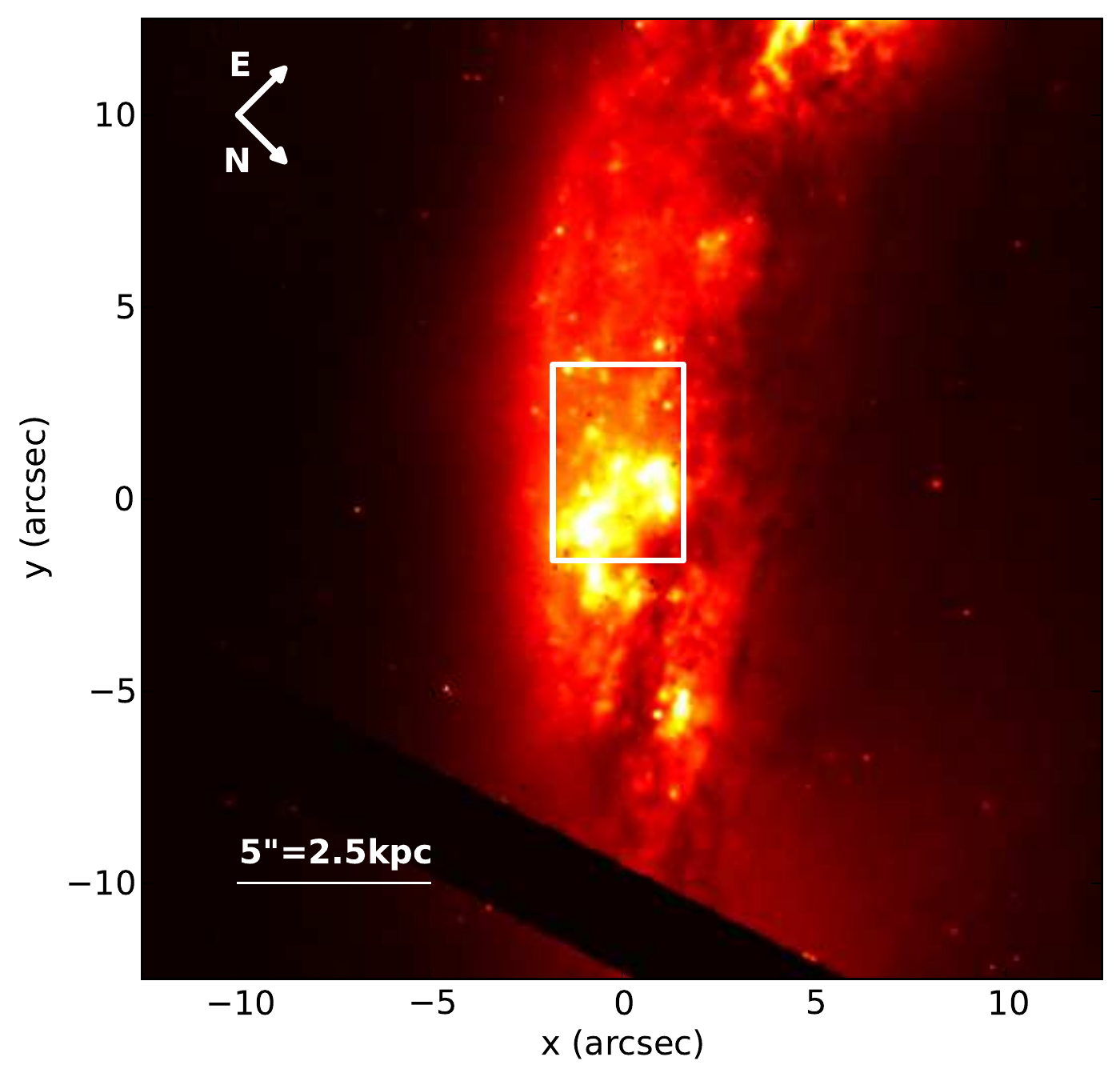}&
\includegraphics[scale=0.56]{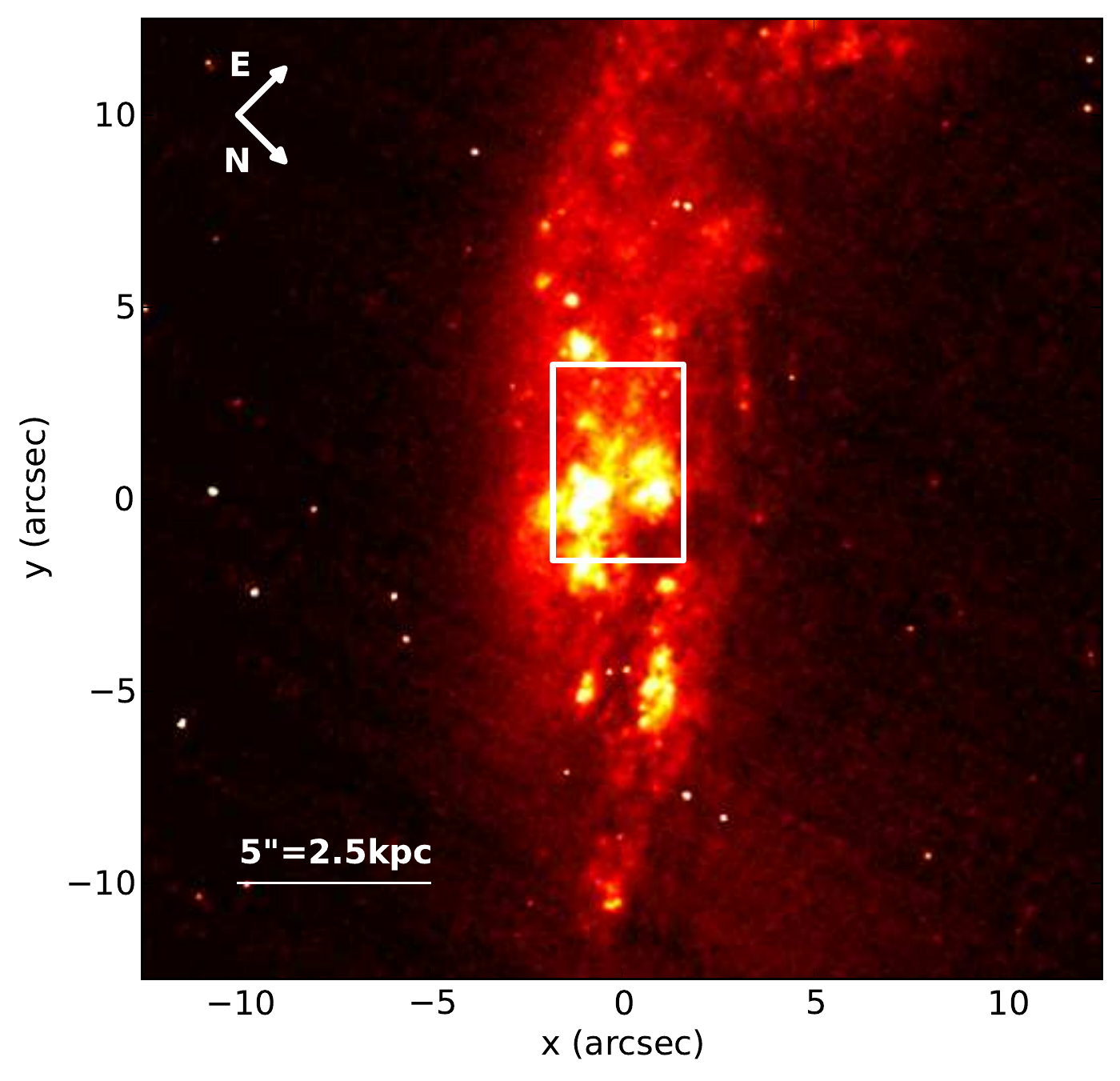}\\
\multicolumn{2}{c}{\includegraphics[scale=0.9,trim=50 365 0 250]{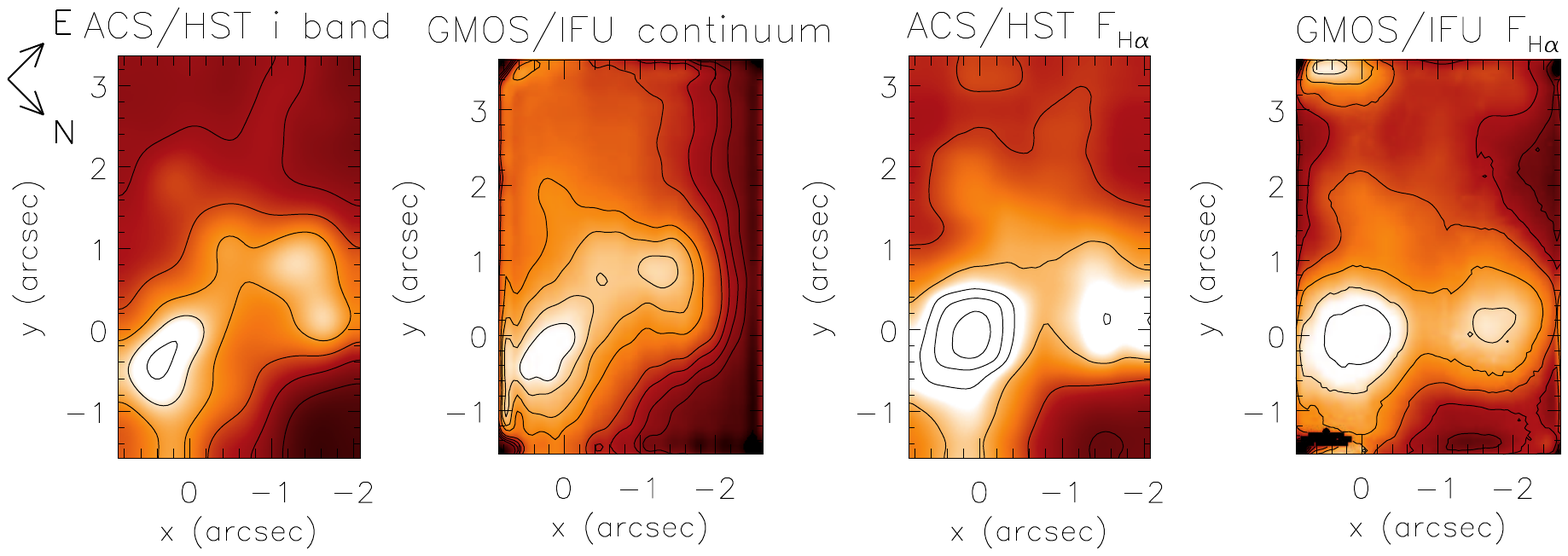}}\\
\end{tabular}
\caption{HST/ACS and reconstructed GMOS/IFU images of IRAS17526+3253. Top-left: large-scale HST/ACS F814W ($i$ band) image. Top-right: large-scale 
continuum free H$\alpha+$[N{\sc ii}] image derived from HST/ACS FR656N and FR914M ramp filter images. 
The boxes represent the FOV shown in the middle panels.
Middle-left: medium-scale image from ACS/HST F814W centered on the northwestern
nucleus that was observed with GMOS/IFU. Middle-right: the same for the HST H$\alpha+$[N{\sc ii}] 
image. The boxes represent the GMOS/IFU field of view. Bottom panels show images
of the GMOS/IFU FOV: from left to right, HST/ACS F814W $i$
band, GMOS/IFU continuum ($\lambda\lambda$ 5500 - 6500\AA), continuum free HST H$\alpha+$[N{\sc ii}],
H$\alpha$ flux map reconstructed from the GMOS/IFU spectroscopy.}
\label{fig:morphology}
\end{figure*}

Two-dimensional spectroscopic observations of IRAS17526+3253 were obtained using the Gemini 
Multi-Object Spectrograph with the Integral Field Unit \citep[GMOS/IFU,][]{Allington2002}. 
The observations were made at the Gemini North telescope on 2013 May, 15-16 UT, as part of 
program GS-2013A-Q-92. We used the IFU 1-slit mode with the B600$+$GG455 grating, which 
encompassed the spectral range $\lambda\lambda\,4855-7730$\,\AA\AA~ and spectral resolution of 3.9\AA, 
covering a field of view (FOV) of roughly 5\farcs1 x  3\farcs4. 
The IFU/FOV was centered on the complex of bright H$\alpha+$[N{\sc ii}] emission associated with the northwestern
galaxy of the interacting pair (figure~\ref{fig:morphology}) and oriented along PA = 221$^{\circ}$. The total on-source integration 
time was 2 hours, taken as six individual exposures of 1200s. 

Data reduction was performed using the sub-packages of {\sc gemini.gmos} {\sc iraf} 
package and followed standard procedures including trimming of the images, bias subtraction,
flat-fielding, wavelength calibration, sky subtraction, and relative flux calibration. We
collapsed the individual data cubes in the spectral range of $\lambda\lambda$5500-6500\AA\AA~  
to create a GMOS/IFU continuum image similar to the HST F814W image. 

Thereafter, we determined the centroid of each dithered image, and then spatially 
shifted and median combined the data cubes using the IRAF \emph{imcombine} task with 
a \emph{sigclip} algorithm to eliminate the remaining cosmic rays and bad pixels. 
In order to suppress high-frequency noise in our data,
we also applied the Butterworth filter\footnote{http://www.exelisvis.com/docs/BANDPASSFILTER.html} 
of 0.2Ny set as upper limits of the pass-through frequency band to the data cube. 

The final cube contains $\sim$ 7000 spectra, each of which corresponds 
to a ``fiber'' of dimensions 0\farcs1$\times$0\farcs1.
The angular resolution of $0\farcs6$ was derived from the FWHM of the spatial profile 
of the standard star. This resolution corresponds to a spatial scale of $\approx300$~pc at the redshift 
of IRAS17526+3253. 

\begin{figure*}
\centering
\begin{tabular}{cccc}
\multicolumn{2}{c}{\includegraphics[scale=0.5]{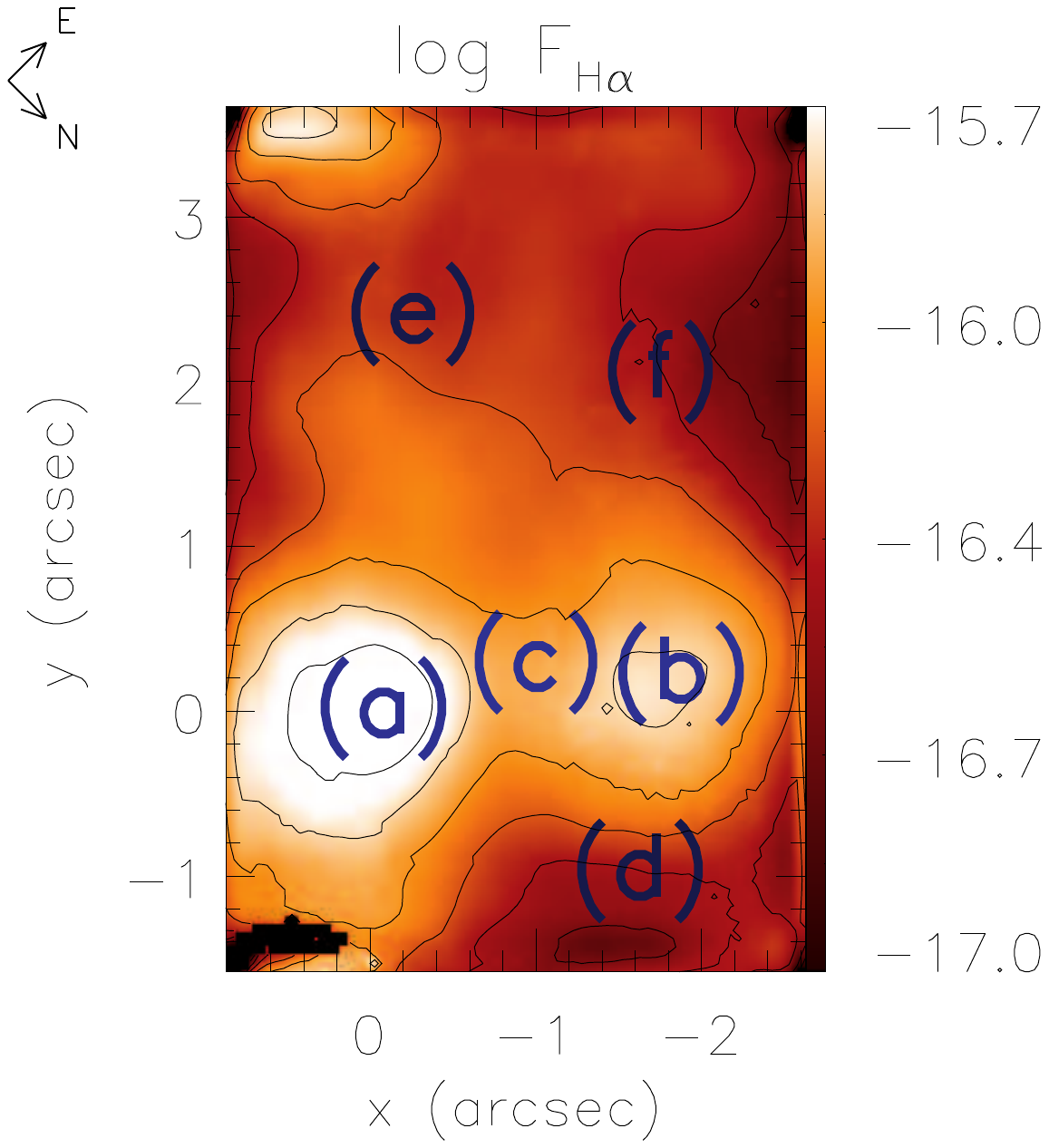}} &
\multicolumn{2}{c}{\includegraphics[scale=0.35]{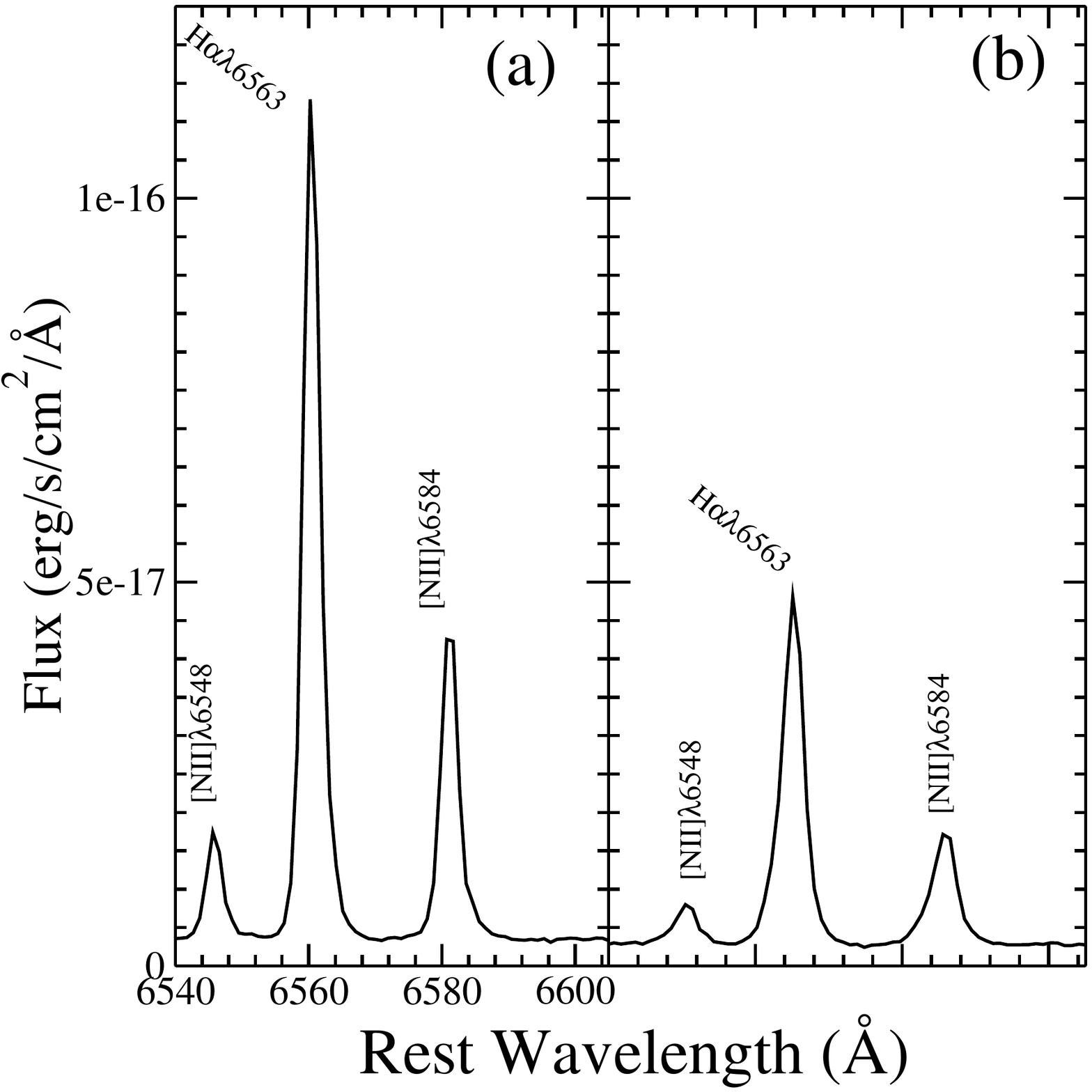}}\\
\multicolumn{4}{c}{\includegraphics[scale=0.45]{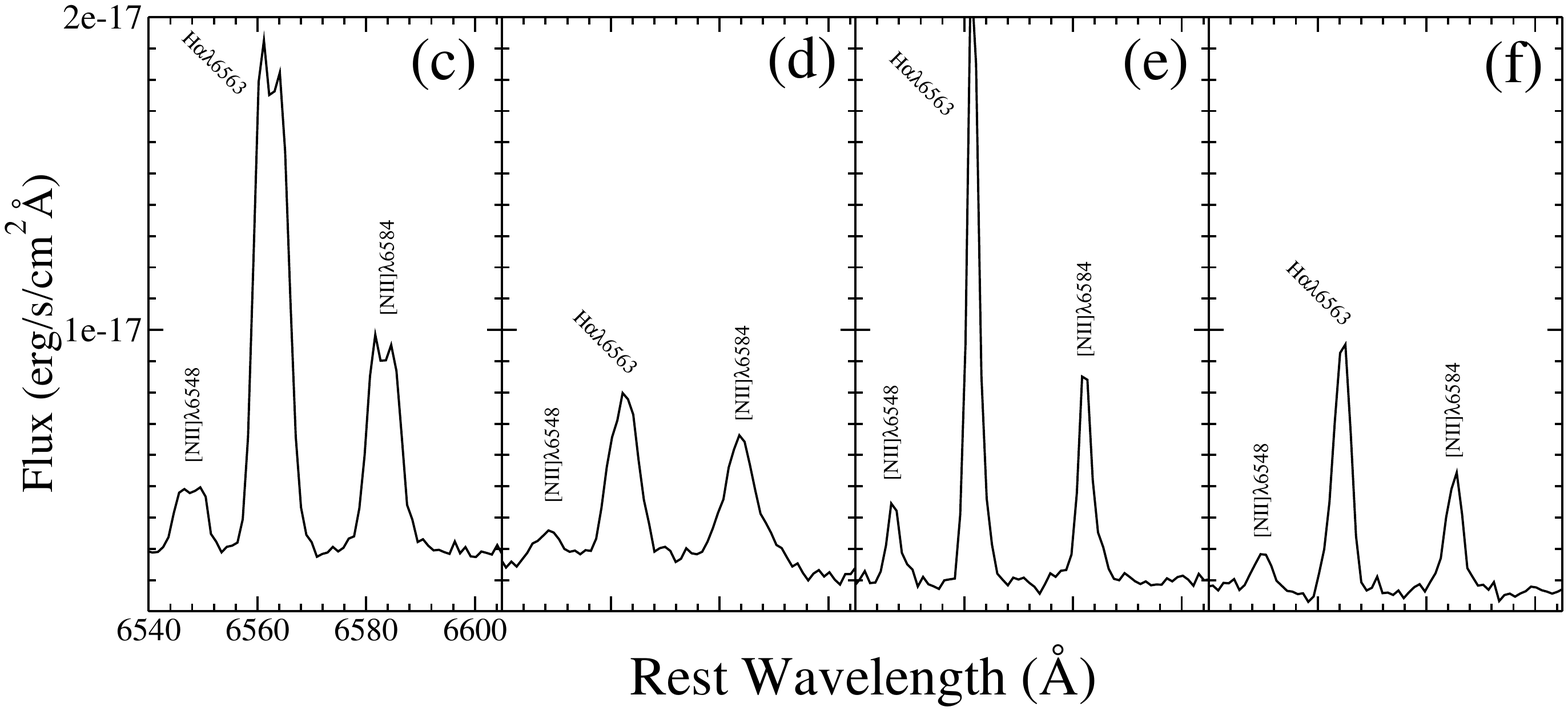}}\\
\end{tabular}
\vspace{-3cm}
\caption{Top-left panel: emission-line map of H$\alpha$ reconstructed from the GMOS/IFU 
spectroscopy and flux unit are shown in logarithmic scale of erg\,s$^{-1}$\,cm$^{2}$\,fiber$^{-1}$.
Top-right panel: spectra corresponding to left and right blobs seem in the
H$\alpha$ image. Bottom-panels: sample of spectra extracted at the positions (c), (d), (e) 
and (f) labeled in the H$\alpha$ flux map.}
\label{fig:espectros}
\end{figure*}

\begin{figure*}
\centering
\includegraphics[scale=0.4]{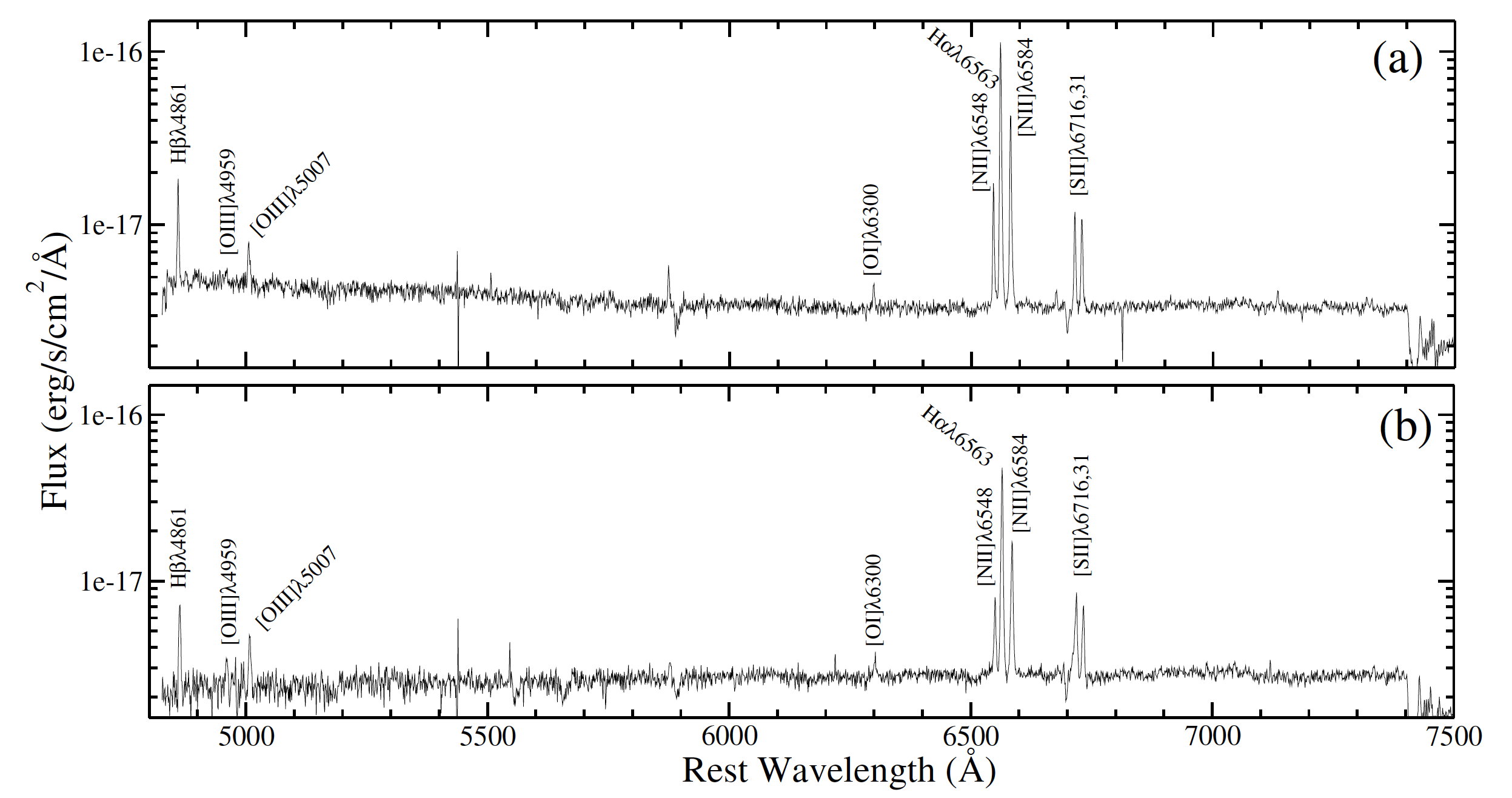}
\caption{The whole observed GMOS/IFU spectral range of NW$_a$ and NW$_b$. The emission 
lines are labeled.}
\label{fig:esp-blobs}
\end{figure*}

In figure \ref{fig:morphology} the HST/ACS broad-band and continuum-subtracted H$\alpha+$[N{\sc ii}] images 
are compared with corresponding continuum and  H$\alpha$  images derived from the GMOS/IFU observations.
The top left and right panels show, respectively, HST i-band and emission line images of the IRAS17526+3253 system.
The middle panels show a region of $30''\times30''$ 
centered at the northwestern nucleus, with the GMOS/IFU FOV being represented by a white box. 
The bottom panels show the HST and corresponding GMOS/IFU images within the IFU FOV. 
The GMOS/IFU continuum image was constructed by collapsing the data cube along the spectral axis between 
5500 and 6500\AA. The H$\alpha$ image was obtained by fitting the emission line profiles in each spaxel 
(see Section \ref{sec:ifu_results}). The HST images were convolved by a Gaussian function with sigma equal 
to 5 pixels ($\sim$0.25'') and spatial registration was carried out by cross-matching these images with the 
corresponding IFU images in order to make the spatial resolution closer to that of the GMOS/IFU data. 
Throughout this paper, we define the origin of the spatial scale to be that of the spaxel in the GMOS/IFU 
H$\alpha$ FOV corresponding to the peak surface brightness of the brightest knot (bottom left in rightmost bottom 
panel of figure \ref{fig:morphology}), whose coordinates are RA: 17$^h$54$^m$29.4$^s$ and DEC: +32$^d$53$^m$12.8$^s$


\section{Results}\label{sec:results}
\subsection{Optical to Near-IR Morphologies}\label{sec:result_opt_nir}

The HST/ACS images in the top row of figure \ref{fig:morphology} clearly reveal a merger system 
in mid-stage with two main galaxy nuclei separated by a projected distance of $\sim$8.8\,kpc. 
The northwestern galaxy appears to present a nearly edge-on aspect, and is crossed by a prominent 
dust lane, suggesting that the northern side is the near side. The disrupted disk of the second 
galaxy (Eastern nucleus) of the interacting pair appears to the merging with that of its companion 
and features a complex system of dust lanes.


The two nuclei are embedded in an elongated irregular envelope, which is interspersed with extensive 
dust lanes and extends over at least 60'' (30\,kpc), with its major axis oriented approximately northwestern--southeastern.
The elongation suggests that we are observing the interaction quite close to the orbital plane. 
The H$\alpha+$[N{\sc ii}] emission image, which is a tracer of young stars via H{\sc ii} regions, 
shows numerous  compact regions of ongoing star formation across the envelope over scales of a few 
10's of kpc. \citet{Baan2006} have shown that the radio emission of these smaller knots in IRAS17526+3253 
is probably associated with SNRs in massive star-forming regions.

The middle panels of figure \ref{fig:morphology} show a region of $30''\times30''$ 
centered on the northwestern nucleus with the GMOS/IFU FOV being represented by a white box. 
The HST/ACS $i$-band continuum image shows a very disturbed morphology with two bright ``blobs'' 
within the IFU field. These features are also prominent in the H$\alpha+$[N{\sc ii}] image, 
indicating that they are star-forming regions. The IFU field was placed to sample these bright 
star-forming regions and it is located just south of the extensive wide dust lane crossing 
the nucleus; the dust lane can be  seen in both the $i$-band and H$\alpha+$[N{\sc ii}] images 
and crosses the northwestern corner of the IFU field. In the bottom row of figure \ref{fig:morphology}, 
the seeing-convolved HST $i$-band and H$\alpha+$[N{\sc ii}] images are compared with corresponding 
images reconstructed from our GMOS/IFU data. The morphological features appearing in the IFU images 
match those seen in the HST well and in particular, the two bright blobs are clearly resolved in the IFU data.

The optical (HST/ACS F814W), near-IR (2MASS K-band) and radio (VLA 1.49\,GHz) morphologies of the whole 
IRAS17526+3253 system are compared in figure~\ref{fig:2mass-radio}. The K-band image is less affected by 
dust extinction and highlights the old stellar population, which can be used as an indication of stellar 
mass distribution of the compact central cores.
The overall morphology in the NIR is similar to that seen at much higher resolution in the $i$-band, 
although the two nuclei are much more prominent relative to the surrounding envelope. 
From Ks band photometry we find that the two nuclei have very similar fluxes, $\sim$12.1 mag,
which translates to a bulge mass of $\sim5.3\times10^9$ M$_{\odot}$, using the scale factor inferred
by \citet{Bell2003}. We also modelled the surface brightness distribution of the 2MASS 
K band image using GALFIT \citep{Peng2002}. It turns out that both nuclei have similar 
S\'ersic indices, indicative of radial brightness profiles consistent with classical bulges (n$_s\approx$ 2.2). 
The main features in the VLA 1.49\,GHz image are two bright, compact 
cores (beam size of 1\farcs35 $\times$ 1\farcs11) that are clearly associated with the two galaxy nuclei, 
as shown by the contours over plotted on the HST $i$-band image in the leftmost two panels.


\subsection{GMOS/IFU data analysis: emission line fitting and kinematics}\label{sec:ifu_results}

This section describes the techniques employed to derive quantitative information from
the GMOS/IFU data cube. To accomplish this, we used customized {\sc profit} routines
written in the IDL\footnote{Interactive Data Language, http://ittvis.com/idl} programming language. 
More information about the {\sc profit} code can be found in \citet{Riffel2010}.

\subsubsection{Single component Gauss-Hermite fits}\label{sec:ifu_ghfit}

The set of spectra in the GMOS/IFU datacube present narrow emission lines, with an average FWHM  of $\sim 5$\AA
The emission line profiles clearly change over the FOV which various regions exhibiting profiles with 
different characteriistics: (i) narrow emission lines displaying a single peak; (ii) emission lines 
characterized by a single peak with asymmetric extended wings due to blueshifted or redshifted components; 
and (iii) line profiles showing a clear double peak. 

We present in figure~\ref{fig:espectros} spectra extracted from spaxels ($0\farcs1\times0\farcs1$) at 
6 locations within the GMOS/IFU FOV, including the peak flux positions of the two bright blobs, labeled as 
(a) and (b) in the figure (top right panels). The other four extractions (labelled as (c), (d), (e) and (f)) 
were selected to illustrate the different emission line profiles and excitation conditions that are present 
within the FOV. In the spectra from regions (e) and (f) as well as those from the blobs (a) and (b), H$\alpha$ 
is much stronger than [N{\sc ii}]$\lambda 6583$ and the profiles are  relatively narrow. In region (c), which 
is located between the two blobs, the lines are broader but double-peaked, suggesting partial blending of 
two components associated with blobs (a) and (b), which present a discontinuity in velocity, as discussed below.
Region (d) on the other hand, exhibits much broader profiles and H$\alpha$ is only slightly stronger than 
[N{\sc ii}]$\lambda 6583$.

Figure~\ref{fig:esp-blobs} shows the entire spectra, covering the whole observed wavelength range, of the
blobs (a) and (b). In addition to H$\alpha$ and the [N{\sc ii}]$\lambda 6548, 83$ lines, H$\beta$,  
[O{\sc iii}]$\lambda 4959, 5007$, [O{\sc i}]$\lambda 6300$ and [S{\sc ii}]$\lambda 6717, 6731$ are also 
present in the spectra. The relative strengths of the lines are consistent with stellar-photoionized H{\sc ii} regions.




As the emission line profiles are clearly non-Gaussian in certain regions, we used a Gauss-Hermite function 
to perform single component fits to each emission lines. The Gauss-Hermite series can be written as
\begin{equation}
\mathcal{GH} = \Psi~~\frac{\alpha(w)}{\sigma}~~[1 + h_3\mathcal{H}_3(w) + h_4\mathcal{H}_4(w)],
\end{equation}
where $w \equiv (\lambda - \lambda_c)/\sigma$ and $\alpha(w) = (1/\sqrt{2\pi})e^{w^2/2}$.

$\Psi$ represents the amplitude of the Gauss-Hermite series, $\lambda_c$ is the central wavelength, 
$h_3$ and $h_4$ are moments of Gauss-Hermite series, which measure respectively, asymmetric and 
symmetric deviations from a pure Gaussian. $\mathcal{H}_3(w)$ and $\mathcal{H}_4(w)$ are Hermite 
polynomials. In order to fit the H$\alpha$+[N\,{\sc ii}] lines we assumed that H$\alpha$ and [N\,{\sc ii}] 
have the same central velocity, while the amplitude of [N\,{\sc ii}]$\lambda$6548 component was 
constrained to be 1/2.96 of that of [N\,{\sc ii}]$\lambda$6583. Similar constrains were applied to 
fit the [O\,{\sc iii}]$\lambda\lambda 4959,5007$ lines. The [S\,{\sc ii}]$\lambda\lambda 6717, 6731$
lines were constrained to have the same central velocity.

\begin{figure*}
\centering
\includegraphics[scale=0.9,trim=0 350 0 70]{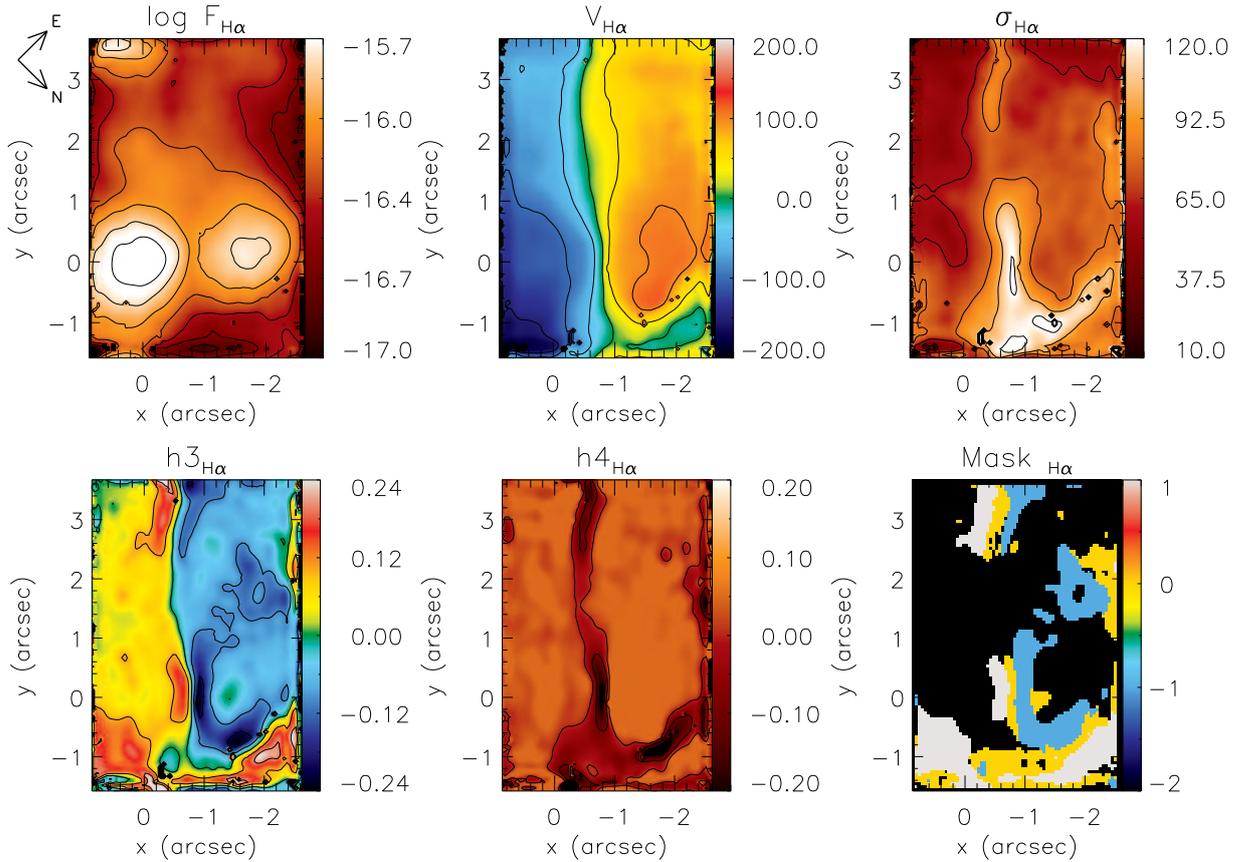}\\
\caption{Results from Gauss-Hermite fits of H$\alpha$ emission line. Top-panels show flux
emission maps (left-panel), velocity field (middle-panel) and velocity dispersion (right-panel) maps.  
Bottom-left and middle panels show Gauss-Hermite h$_3$ and h$_4$ moments. Bottom-right panel
presents masked regions that received values from two-gaussian fits. Flux units are shown in 
logarithmic scale of erg\,s$^{-1}$\,cm$^{2}$\,fiber$^{-1}$.}
\label{fig:gauss-hermite}
\end{figure*}

The results of the emission line fitting are shown in figure~\ref{fig:gauss-hermite}. 
The values of $h_3$ and $h_4$ vary from $-0.2$ to 0.2. Visual inspection of the fits 
suggests that emission lines with blue wings show values of $h_3 < -0.1$, while red wings 
appear in regions with $h_3 > 0.1$. Values of $h_4 < -0.1$ select regions where the line 
profiles are broader (not as sharply peaked) than an equivalent Gaussian of the same amplitude, 
including profiles that have noticeable double peaks. The values of $h_3$ and $h_4$ recovered from 
the fits therefore allow extended asymmetric wings and double peaked profiles to be automatically 
identified. We used these values to generate a mask which flags spaxels that required more than 
one Gaussian to fit the observed emission lines. The mask image is also shown in figure~\ref{fig:gauss-hermite}.

The H$\alpha$ flux map obtained by Gauss-Hermite fitting highlights the presence of the two bright 
blobs that are the most prominent features in the IFU FOV, as already seen in the HST $i$-band and 
H$\alpha+$[N{\sc ii}] emission line images (see figure~\ref{fig:morphology}). A smaller and somewhat 
fainter third blob is located near the  southern corner of the FOV, which appears to be associated with 
a bright knot seen in the HST H$\alpha+$[N{\sc ii}] image just beyond the southeastern border of the IFU field 
(see figure \ref{fig:morphology}). The flux distributions of the other emission lines (with the exception 
of [O\,{\sc i}]$\lambda\lambda$6300) show  features similar to those observed in the H$\alpha$ map, 
but with lower S/N, and thus we do not present them in this section. 

It is immediately clear from visual inspection of the data cube that the emission lines of the 
brightest blob (northwestern, hereafter NW$_a$) are blueshifted with respect to the line emission at the 
center of GMOS/IFU FOV, while the northeastern blob (hereafter NW$_b$) appears to have redshifted emission 
lines. We therefore determined the systemic velocity for the IFU field  by averaging the velocities of two 
bright blobs ($v_{sys}$ = 7580 km s$^{-1}$) within apertures of $0.75''$ centered on the flux peaks. 
Maps of the centroid velocity, corrected for this systemic velocity, and the velocity dispersion, corrected 
for the instrumental resolution, are shown in figure~\ref{fig:gauss-hermite}.

The velocity map shows blueshifts dominating the SW side of the FOV, while redshifts dominate the NE side. 
These regions are very clearly segregated, with a sharp transition between the blue and redshifted sides.
This velocity field does not show the ``spider'' diagram typical of disk rotation. Moreover, 
given the morphology and apparent orientation of the northwestern galaxy, we would expect the position angle 
of the line of nodes to be approximately northwestern-southeastern, nearly orthogonal to the direction one would infer from the 
velocity map. It is also notable that the two regions, or the boundary between them, are clearly distinguishable 
in the $\sigma$, $h_3$ and $h_4$ maps. Thus, $h_3$ shows that the emission lines in the SW (blueshifted) region 
are characterized by red asymmetries, whereas in the NE (redshifted) region the lines are typically blue asymmetric. 
The $\sigma$ and $h_4$ maps both show narrow features that coincide with the sharp boundary between the blue and 
redshifted regions, indicating that the lines are broader and more flat-topped along the boundary. Indeed, this is 
where we observe double-peaked profiles as exemplified by the region (d) spectrum shown in figure~\ref{fig:espectros}. 
Taken together, the kinematic parameter maps suggest that we are observing two distinct, partially superimposed
kinematic components which have a relative velocity $\sim 200$km\,s$^{-1}$.
 
There is also evidence for a third kinematic component, which is traced by the region of increased velocity 
dispersion close to the northwestern edge of the FOV, with a spur running NE. Similar structures are seen in the $h_3$ 
and $h_4$ maps, indicating red asymmetric, broader lines.

\subsubsection{Two component Gaussian fits.}

As described above, the results of the single-component Gauss-Hermite fits have been used to identify spaxels
where the emission lines cannot be well-fitted with a single Gaussian profile (see figure \ref{fig:gauss-hermite}). 
These include spaxels where the lines have a single peak but extended or asymmetric  wings, as well as those 
where the lines exhibit double peaks. As these profile shapes may result from blending of kinematic components 
arising from distinct kinematic components, we have used two Gaussian components to fit the line profiles in those spaxels
that have $-0.1 < h_3 < 0.1$ and $h_4 < -0.1$. The first criteria track spaxels that have emission line
with extended wings, once that second one identifies spaxels with doble peaks of the emission profile.
These regions are indicated with  values between $-1$ and 1 in the mask image shown in figure \ref{fig:gauss-hermite}.
A single Gaussian component was used to fit the line profiles over the remainder of the FoV
The same constraints as used in the Gauss-Hermite fits (Section~\ref{sec:ifu_ghfit}) were applied to each 
gaussian component in the fits to the H$\alpha$+[N\,{\sc ii}], [O\,{\sc iii}]$\lambda\lambda 4959,5007$ and 
[S\,{\sc ii}]$\lambda\lambda 6717, 6731$  line groups.

\begin{figure*}
\centering
\includegraphics[scale=0.52]{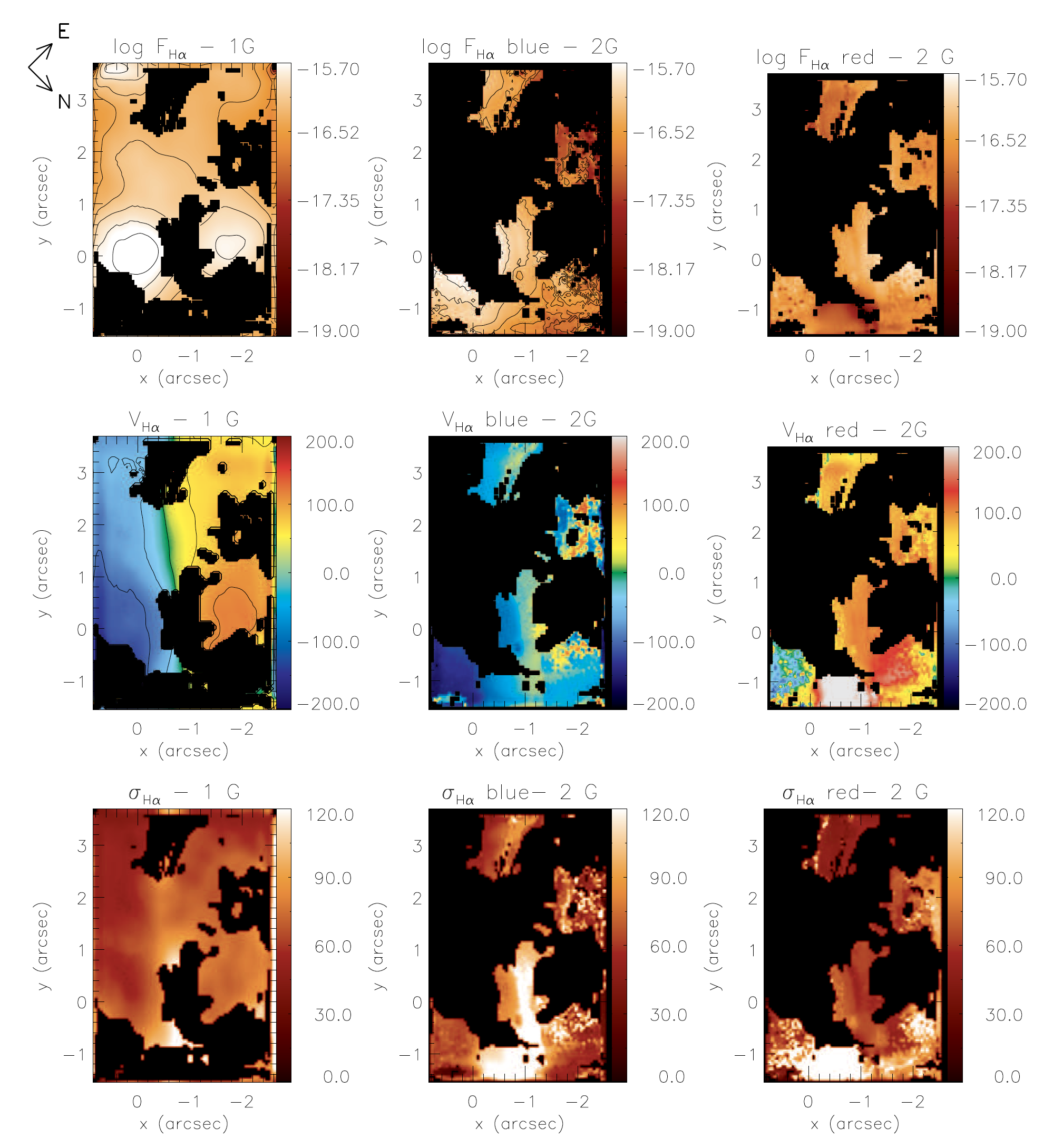}\\
\caption{Top-panels: emission line maps of H$\alpha$. Left-top-panel is emission flux taken
from one Gaussian fitting. Middle and right top-panels are respectively the blueshifted and 
redshifted components derived from two Gaussian fitting of H$\alpha$ emission. Middle-panels
are the same for velocity maps. Velocity dispersion are showed in the bottom-panels. Flux 
units are shown in logarithmic scale of erg\,s$^{-1}$\,cm$^{2}$\,fiber$^{-1}$.}
\label{fig:maps-mask}
\end{figure*}

The flux distribution, velocity and velocity dispersion (after correction for instrumental resolution) maps derived 
from the single and double gaussian fits are presented for the H$\alpha$ emission line in figure \ref{fig:maps-mask}.
It can be seen from the flux distributions that the region in which single gaussian fits were used includes the two 
prominent flux peaks previously identified as NW$_a$ and NW$_b$, whereas the regions where the profiles exhibit 
splitting or extended wings and where two gaussian components were used, mainly lie between the blobs, or border them 
along the northwestern edge of the FOV.

Not surprisingly, the velocity field within the region where the lines were fitted with a single Gaussian component 
is very similar to that obtained from the single Gauss-Hermite series, exhibiting the same pattern of blueshifts 
on the SW side of the FOV, and redshifts on the NE side. The velocities obtained from the blue and red-shifted components 
of the two-component Gaussian fits are generally comparable with those obtained from the single component fits in the blue- 
and red-shifted sides of the FOV. The main exception is a region close the northwestern border of  the FOV, where the red component 
has a much higher velocity ($\sim 200$\,km\,s$^{-1}$) than is seen elsewhere. This location also coincides with a high 
velocity dispersion in both the red and blue components. However, the blue component in this region does not show a 
correspondingly large blueshift and its velocity in fact is quite comparable with the blueshifted velocities over the 
SW side of the FOV.
 

\begin{figure*}
 \centering
   \includegraphics[scale=0.9,trim=50 360 0 0]{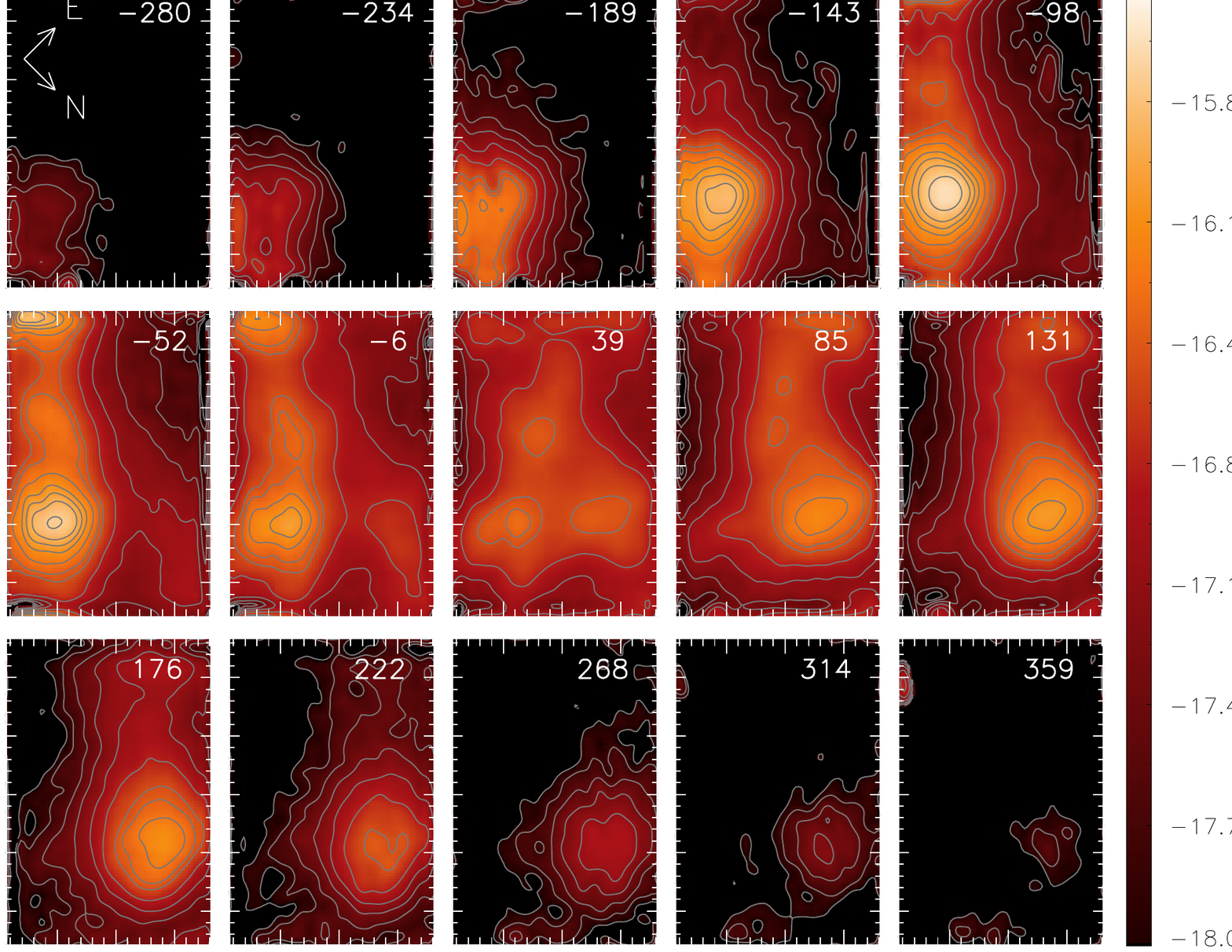}
\caption{Observed radial velocity channel maps of the H$\alpha$ emission 
line are plotted with a logarithmic scale. The radial velocity range from 
-337 to +303 km s$^{-1}$ in a of 46 km s$^{-1}$. The velocity of each channel is
shown in the top-right corner in the corresponding panel. The intensity scale
is displayed at the right side.}
\label{fig:channel-map}
\end{figure*}

In figure \ref{fig:channel-map} we present velocity channel maps extracted along the H$\alpha$ line
profile, using velocity bins of 45 km s$^{-1}$. The distribution of H$\alpha$ emission in these velocity slices 
also suggests that there are two distinct but partially overlapping kinematic components. 
Approximately half of the FOV on the SW side is covered by predominately blueshifted emission which includes the 
prominent flux peak, NW$_a$, 
as well the fainter blob at the south corner of the
FOV, which appears to be connected to NW$_a$ by a ridge of emission. These structures can be seen at velocities ranging from  $\sim -200$ to $+70$\,km s$^{-1}$ 
 at NW$_a$. The NE half of the FOV is dominated the redshifted emission. The main feature is NW$_b$, but there is also a ridge of emission extending to the SE, with these structures
 spanning the velocity range $\sim 70-300$\,km s$^{-1}$.

To gain further insights into the correlations between the spatial distribution of the line emitting gas and its velocity field, 
we applied Principal Components Analysis (PCA) to the  IFU data cube. 
PCA is a statistical technique that has recently found various applications in astrophysics and in particular has been extensively applied to
analyse IFU observations  \citep[PCA, e.g.][]{Murtagh1987,Steiner2009,Ricci2011}. 
This tool transforms the system of correlated coordinates into a
system of orthogonal coordinates, and eigenvectors, ordered by principal components of decreasing variance. The
projection of the data onto these coordinates produces images called tomograms, while eigenvectors correspond
to the eigenspectra. The combined analysis of both can be used to identify distinct components 
in the IFU data, which facilitates physical interpretation.

\begin{figure}
\centering
\begin{tabular}{ll}
\includegraphics[scale=0.3]{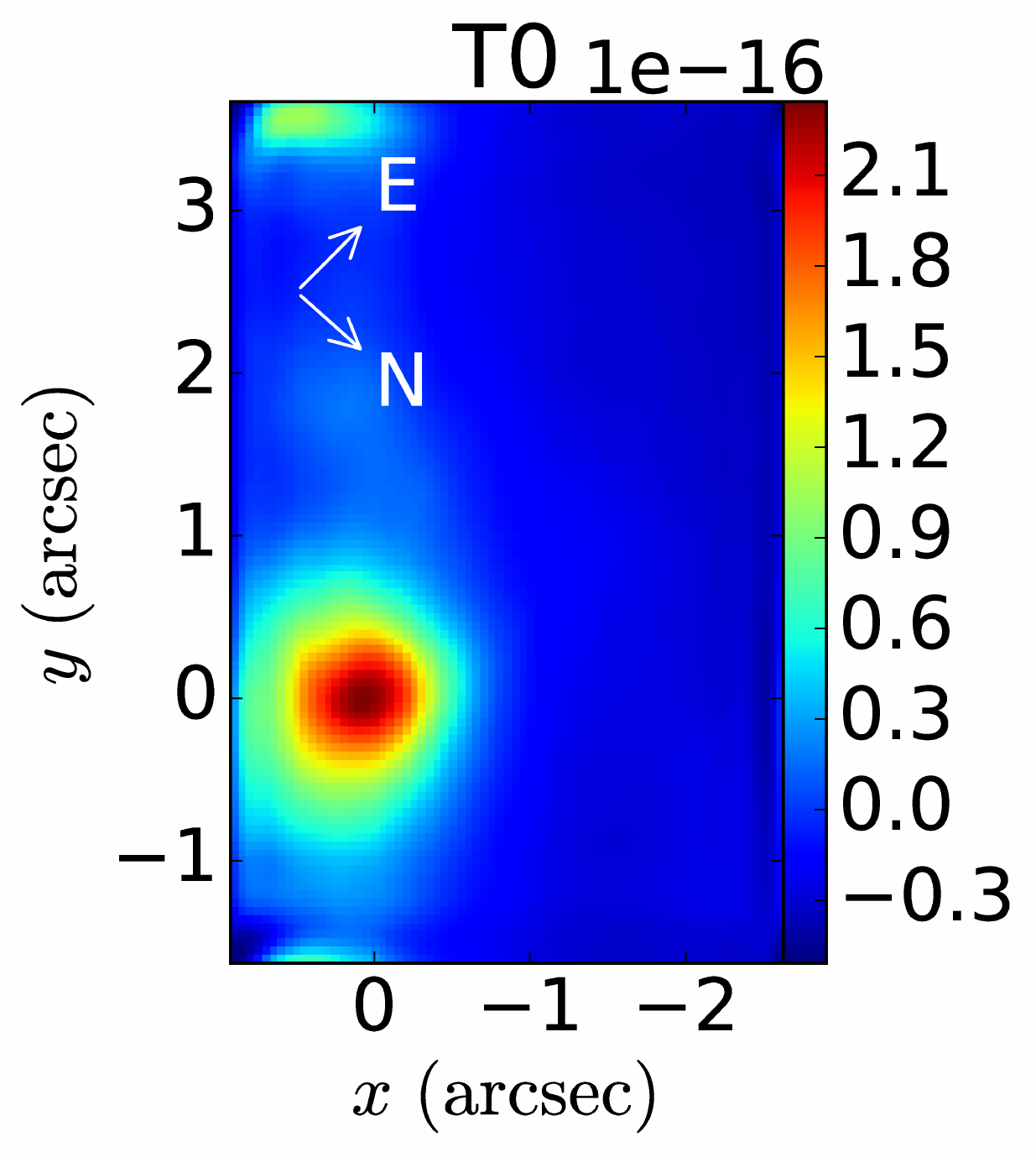}&
\includegraphics[scale=0.28]{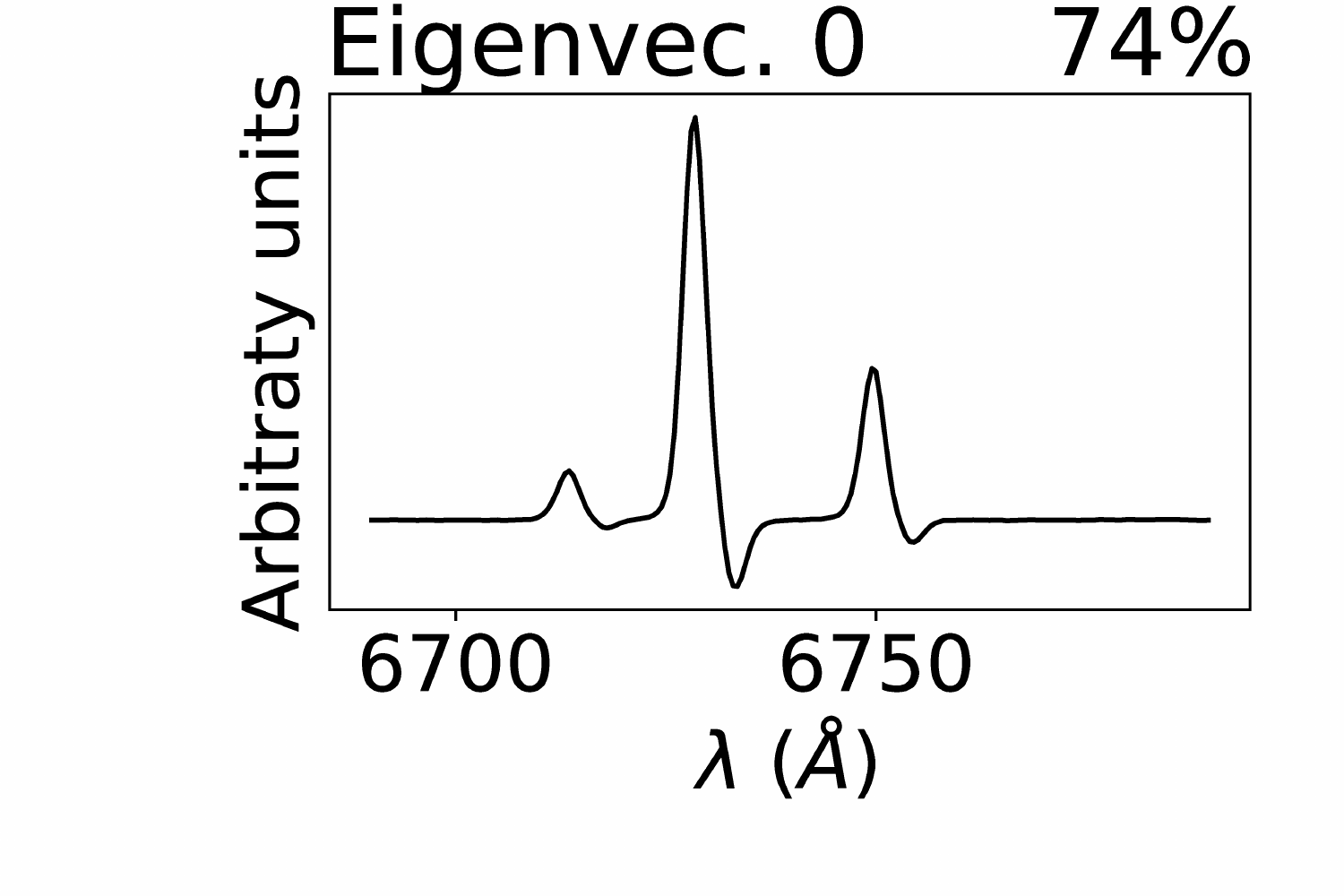}\\
\includegraphics[scale=0.3]{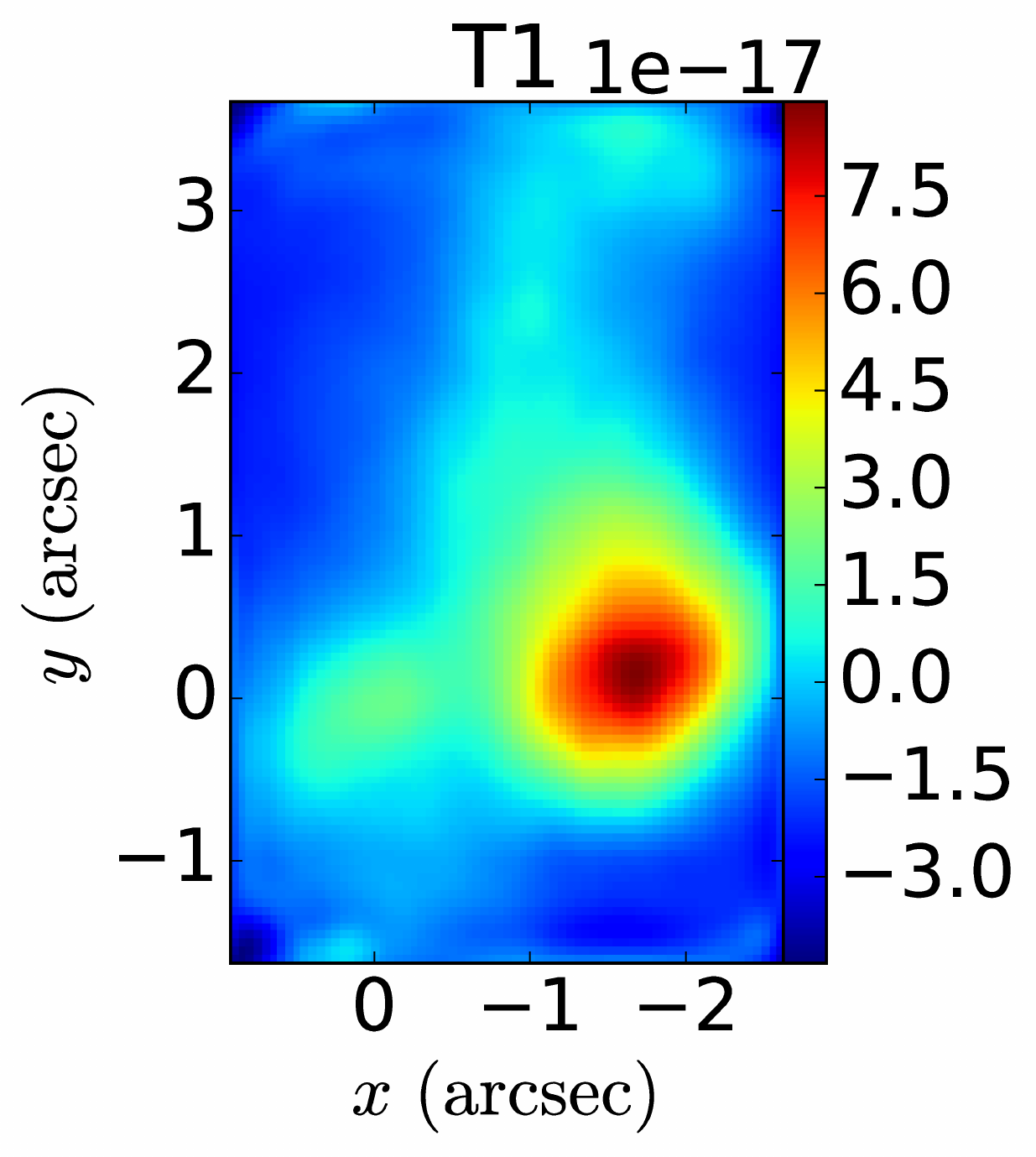}&
\includegraphics[scale=0.28]{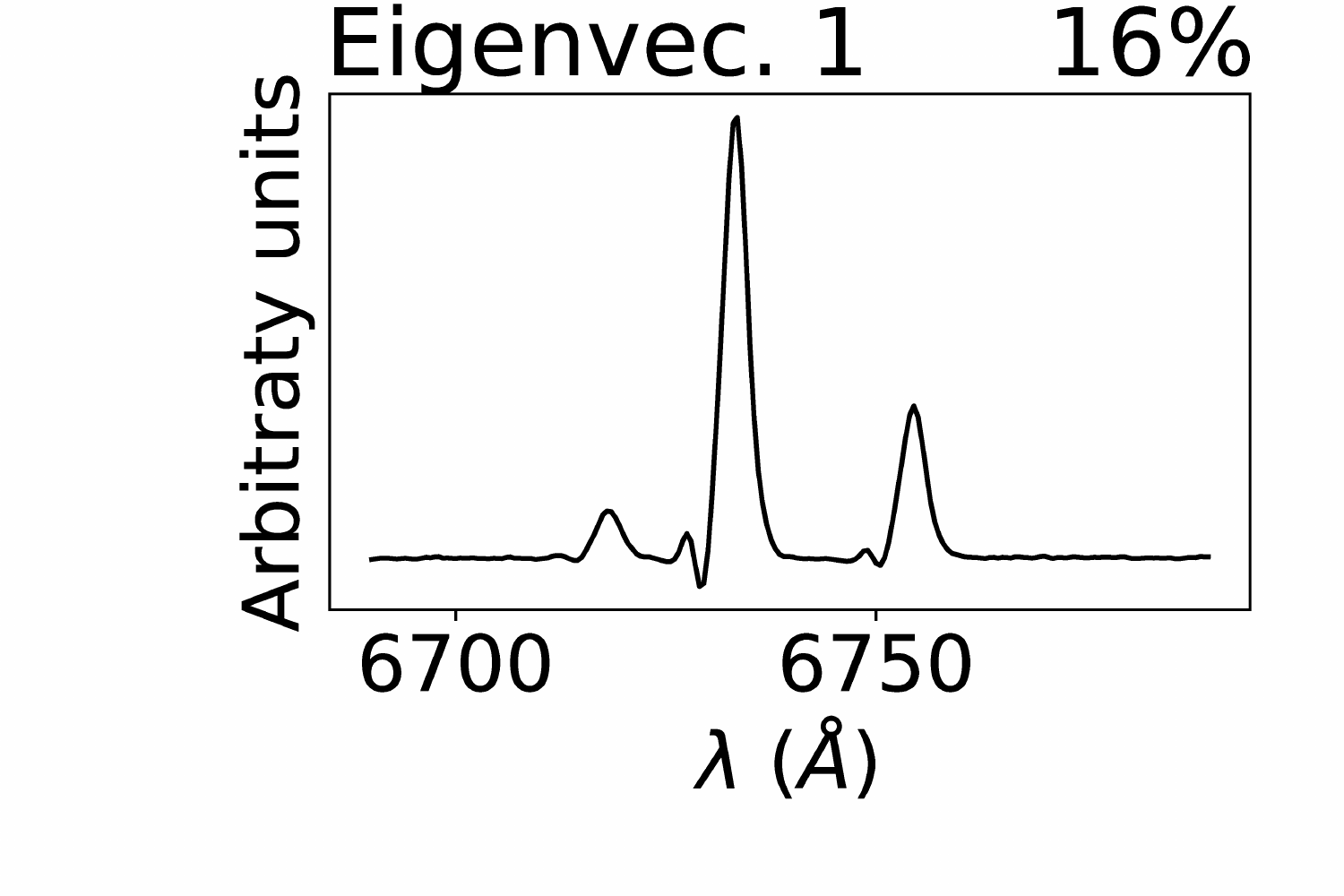}\\
\includegraphics[scale=0.3]{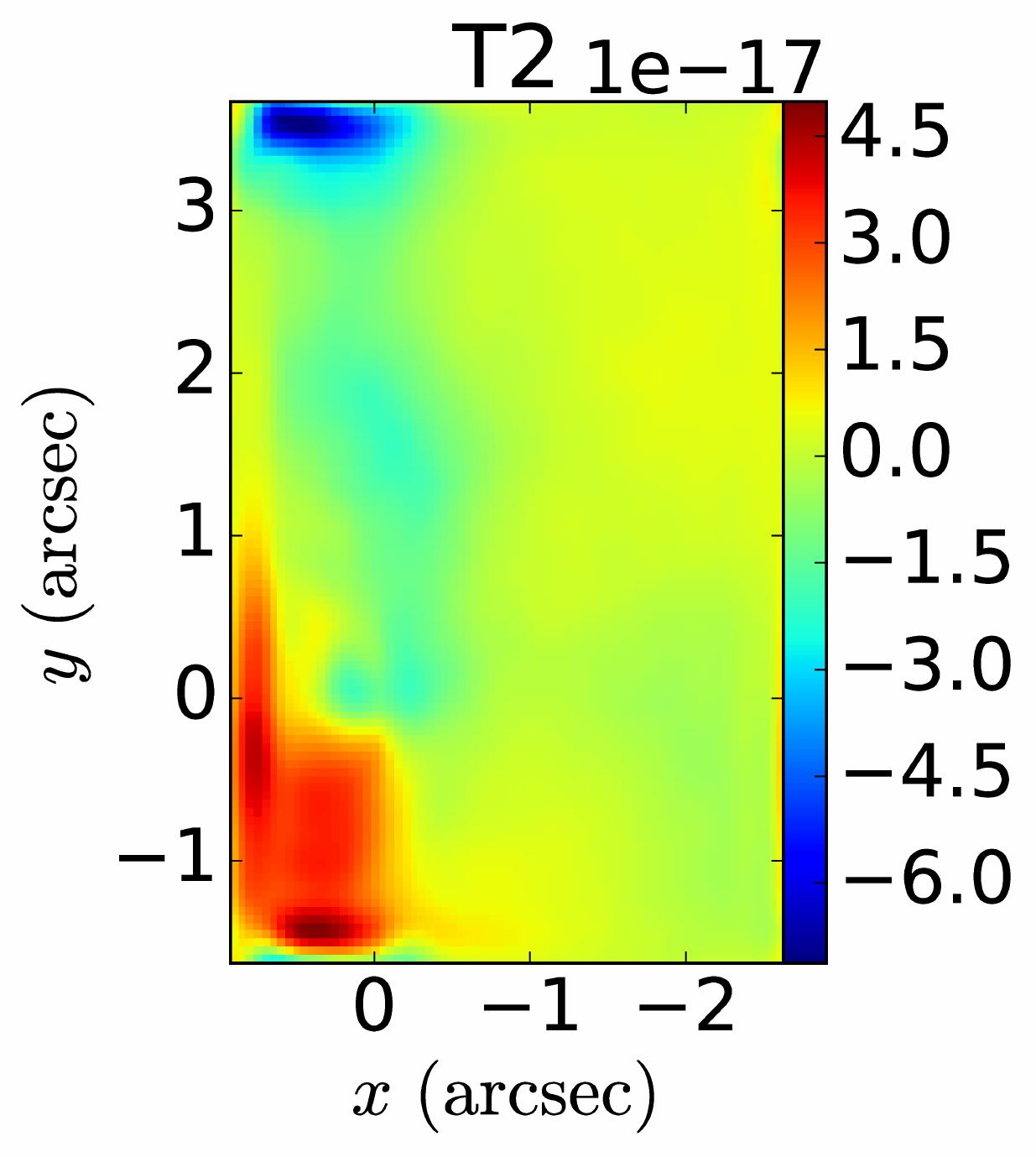}&
\includegraphics[scale=0.28]{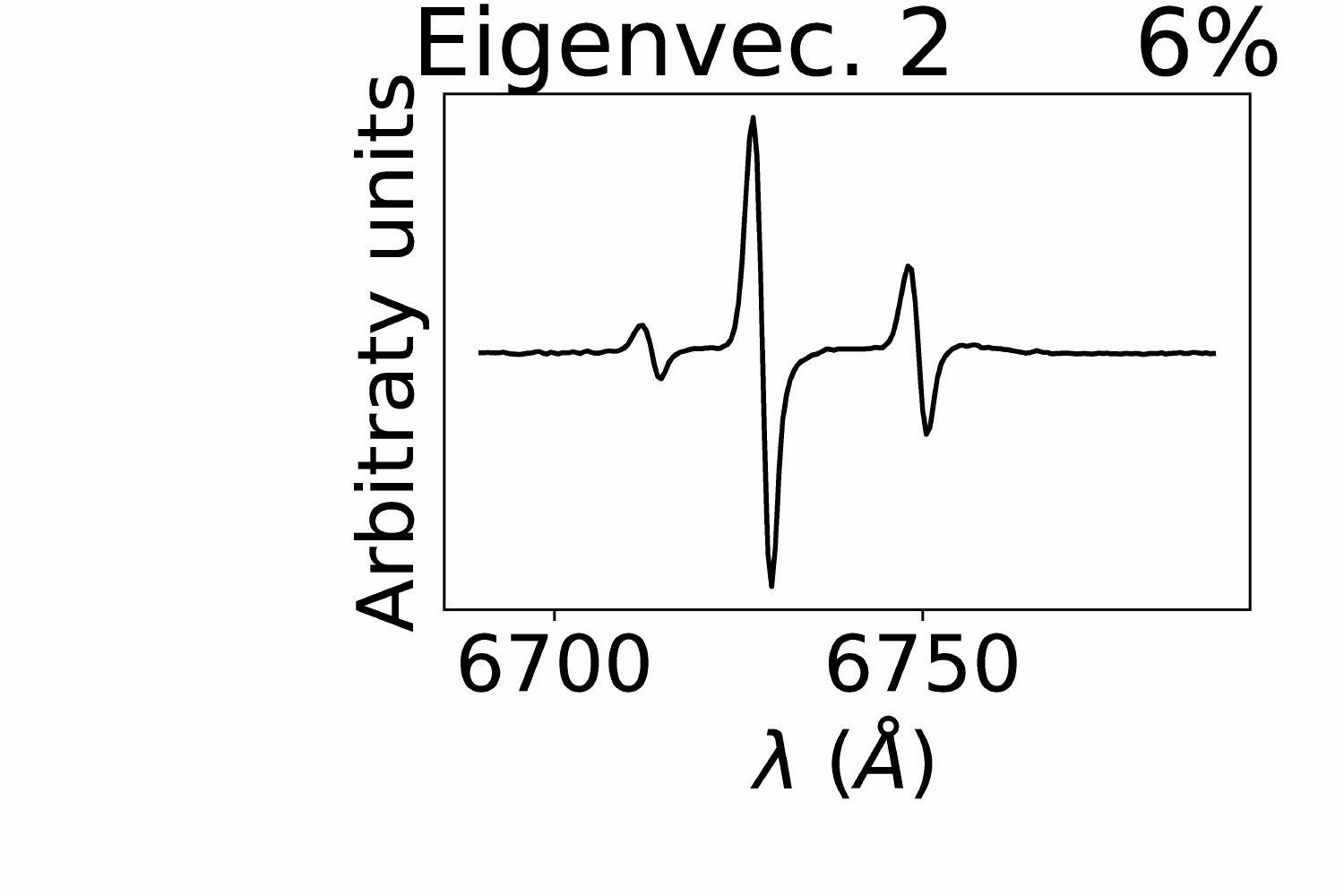}\\
\includegraphics[scale=0.3]{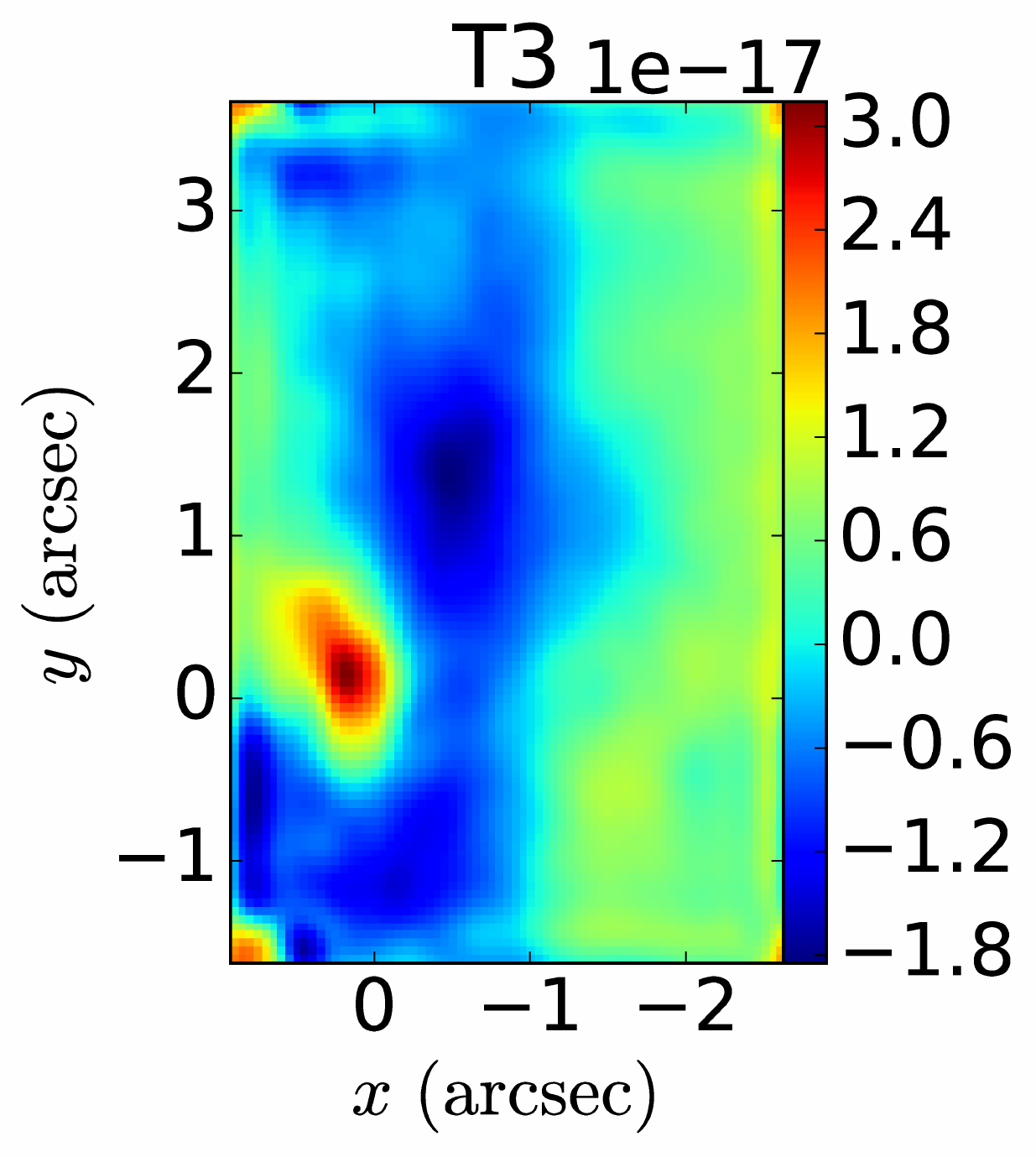}&
\includegraphics[scale=0.28]{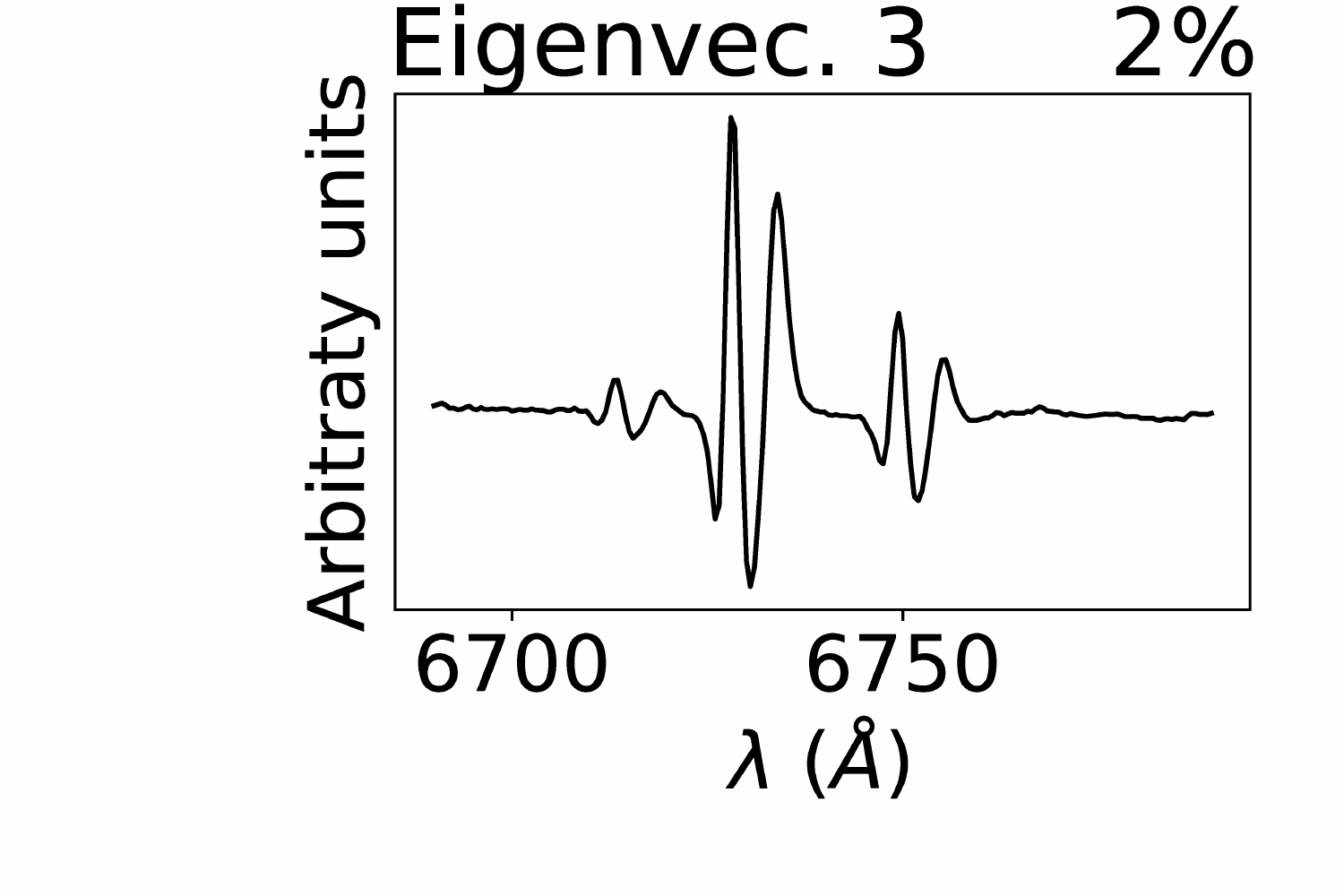}\\
\end{tabular}
\caption{PCA derived from GMOS/IFU data. From top to bottom: tomograms (left-panels)
and eigenspectra (right-panels) corresponding to first four principal components of 
the data cube of IRAS17526+3253.}
\label{pca-tomo}
\end{figure}

The results of our PCA analysis of the IRAS17526+3253 IFU data are shown in figure \ref{pca-tomo}. 
The derived tomograms and their respective eigenspectra are shown in the left and right columns, respectively. The first 
eigenspectrum, which accounts for $\sim 74\%$ of the data cube variance, corresponds to the blueshifted emission and the 
bright NW$_a$ in the SW side of the FOV. The second, which accounts for $\sim 16\%$ of the variance, is 
mainly associated with the redshifted emission including NW$_b$, a weaker structure extending
in the direction of NW$_a$ and a diffuse feature extending to the SE. 
The $3^{rd}$ and $4^{th}$ eigenspectra account for the remaining $\sim 8\%$ of the variance in the data cube and 
exhibit features associated, for example, with the peak of NW$_a$, the knot at the S corner of the FOV and the boundary between the red and blueshifted regions.  
Overall, the structures revealed by the PCA analysis exhibit a very close correspondence with the features seen in the velocity channel maps.

Together with the channel maps, the results of the PCA analysis support the interpretation of the parameter 
maps derived from line-fitting, that the SW and NE sides of the FOV sample kinematically independent systems, 
which are respectively blue- and redshifted relative to the adopted systemic velocity. Accordingly, we have 
constructed separate maps of these regions for selected emission  lines  by combining the velocities, velocity 
dispersions and fluxes from the single component Gaussian fit with the corresponding values for the blue and 
redshifted components, respectively, of the double gaussian fits. 

\begin{figure}
   \includegraphics[scale=0.9,trim=60 360 0 0]{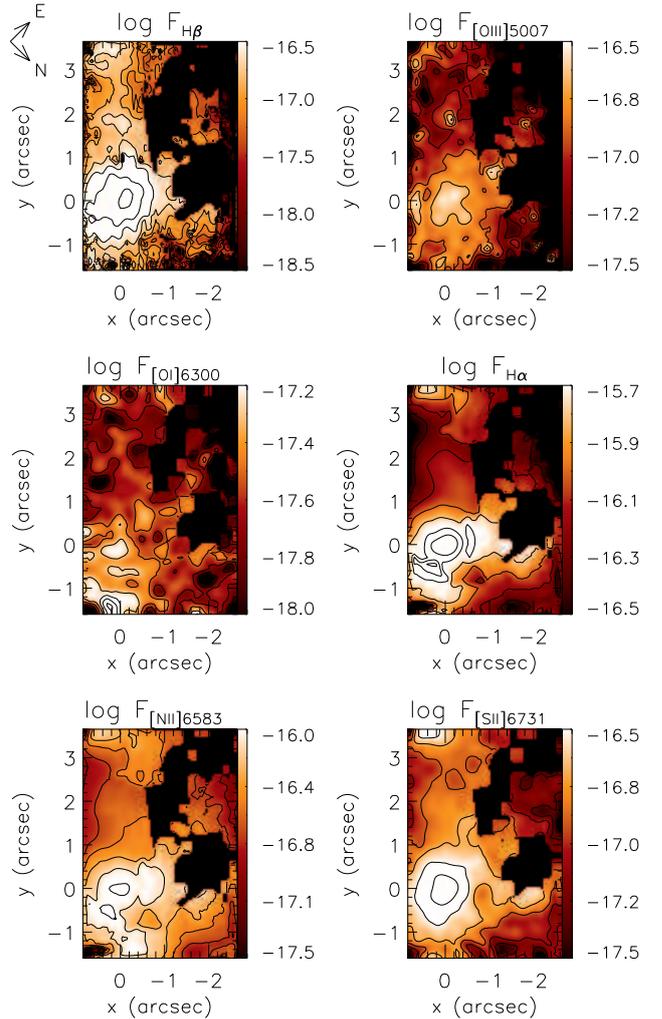}\\
\caption{Total flux maps of the NW$_a$. Flux unit are shown in logarithmic scale of erg\,s$^{-1}$\,cm$^{2}$\,fiber$^{-1}$.
Black regions demonstrate masked location corresponding to the NW$_b$ and/or lower signal-to-noise ratio ($S/N < 3$).}
\label{fig:flux-bloba}
\end{figure}

The flux maps for the H$\beta$, [O\,{\sc iii}]$\lambda\lambda$5007, [O\,{\sc i}]$\lambda\lambda$6300, 
[N\,{\sc ii}]$ \lambda\lambda$6584, H$\alpha$, and [S\,{\sc ii}]$\lambda\lambda$6731 
emission lines are presented for the blueshifted region in figure \ref{fig:flux-bloba}. 
With the exception of [O\,{\sc i}]$\lambda\lambda$6300, the line fluxes show similar distributions,
the main feature being NW$_a$ with an angular size of roughly $0\farcs75$ ($\approx400$pc). The fainter blob 
some $3\farcs5$ to the SE is also seen in all lines, except for [O\,{\sc iii}]$\lambda\lambda$5007 (probably due 
to the lower SNR). The  [O\,{\sc i}]$\lambda$6300 line is relatively weak over most of the FOV (SNR$\approx 5$), 
but it is notable that its flux peak occurs near the northwestern edge of the field, offset by about $1\farcs5$ to the North of NW$_a$.

\begin{figure}
   \includegraphics[scale=0.9,trim=60 360  0 0]{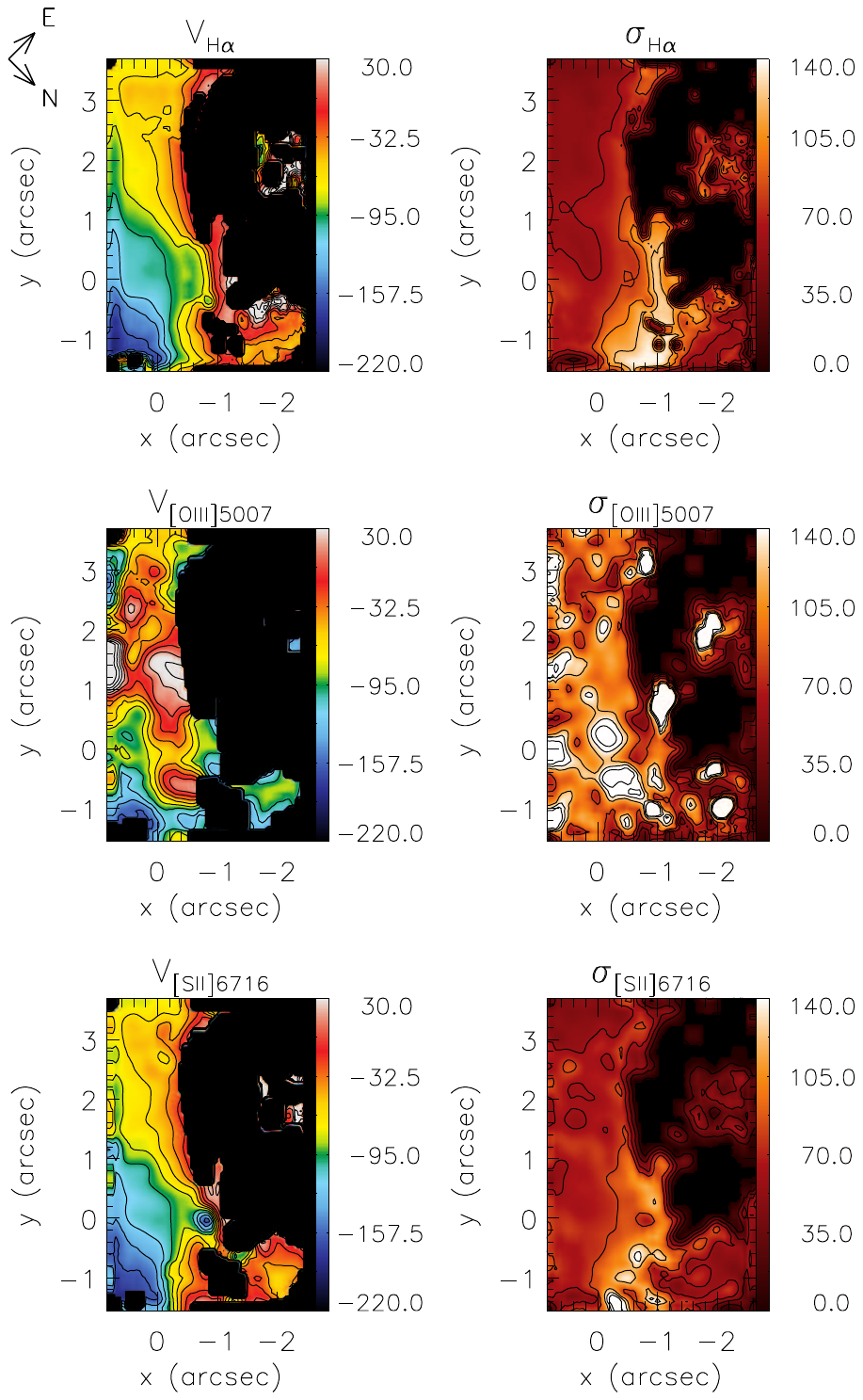}\\
\caption{Centroid velocity and velocity dispersion maps for some emission 
lines of the NW$_a$. Units are in km s$^{-1}$. Black regions demonstrate masked location corresponding 
to the NW$_b$ and/or lower signal-to-noise ratio ($S/N < 3$).}
\label{fig:vel-bloba}
\end{figure}

Figure \ref{fig:vel-bloba} shows the 
centroid velocity (left-panels) and velocity dispersion (right panels) maps for the H$\alpha$, [O\,{\sc iii}]$\lambda\lambda$5007 and 
[S\,{\sc ii}]$\lambda\lambda$6716 lines in the blueshifted region, after subtraction of the systemic velocity ($v_{sys}$ = 7580 km s$^{-1}$) 
and correction for instrumental resolution, respectively. As the corresponding maps for [N\,{\sc ii}]$\lambda$6584 are similar to those 
of H$\alpha$ line, they are not shown. As a whole, the gas in the southwestern half of the FOV exhibits a northwestern-southeastern velocity gradient, with the largest blueshifts ($\sim -150$\,km s$^{-1}$) 
occurring in a region just south of NW$_a$. Blob NW$_a$ itself has a velocity ($\sim -100$\,km s$^{-1}$), while the diffuse emission has velocities ($\sim -50$\,km s$^{-1}$).
The highest velocity dispersions ($\gtrsim 100$\,km s$^{-1}$) occur in a localized region near the northwestern border of the FOV, and roughly centered between the two bright blobs.
A ridge of higher velocity dispersion also extends southeastern along the border between the blue- and redshifted regions. However, outside these localized enhancements, the velocity dispersion is fairly uniform for the remainder of the blueshifted region, with a typical value $\sim 50$\,km s$^{-1}$ (the small-scale structure in the [O\,{\sc iii}]$\lambda 5007$ map is due to the low SNR ratio). In particular, neither NW$_a$ or the secondary knot at the S corner of the FOV correspond to obvious features in the velocity dispersion map.

\begin{figure}
\centering
   \includegraphics[scale=0.9,trim=60 360 0 0]{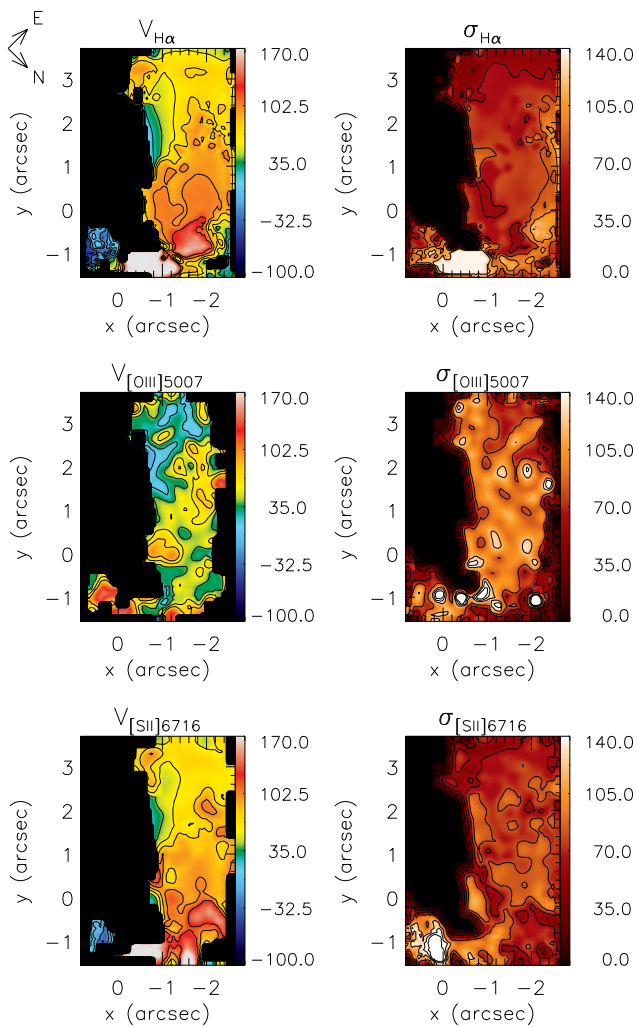}\\
\caption{Examples of the centroid velocity and velocity dispersion maps for some emission 
lines of the NW$_b$. Units are in km s$^{-1}$. Black regions demonstrate masked location corresponding 
to the NW$_b$ and/or lower signal-to-noise ratio ($S/N < 3$).}
\label{fig:vel-blobb}
\end{figure}

The emission-line flux, centroid velocity and velocity dispersion maps for the redshifted region are
presented in figures \ref{fig:flux-blobb} and \ref{fig:vel-blobb}. 
The main emission component in most lines is the component identified as NW$_b$, which has a similar angular size to that of NW$_a$ ($0\farcs75$, and $\approx400$pc).
However, just as in the corresponding flux map for the blueshifted region (figure \ref{fig:flux-blobb}), the strongest [O\,{\sc i}]$\lambda$6300 emission is located close to the northwestern edge of the FOV, bordering NW$_a$ to the N.  To a lesser extent this feature is also seen in [N\,{\sc ii}]$ \lambda\lambda$6584. 

The velocity field of the redshifted region shows a similar gradient to that of the blueshifted region (\ref{fig:vel-blobb}): the highest velocities ($\sim +150$\,km s$^{-1}$) occur just northwestern of NW$_a$, which itself has a velocity $\sim +100$\,km s$^{-1}$, whereas the diffuse emission to the SE has a typical velocity $\sim +50$\,km s$^{-1}$. However, there is also a distinct region, spatially coincident with the region where the  [O\,{\sc i}]$\lambda$6300 emission peaks, which  clearly exhibits a high velocity ($\sim +200$\,km s$^{-1}$) and velocity dispersion ($\sim 150$\,km s$^{-1}$) in H$\alpha$, and possibly the other lines. The velocity dispersion is also generally enhanced ($\sim +200$\,km s$^{-1}$) between the northwestern edge of the FOV and blobs NW$_a$ and NW$_b$. It is worth noting that this side of the IFU FOV borders the radio source associated with the nucleus of the northwestern galaxy (see figure \ref{fig:2mass-radio}).
 Over the rest of the redshifted region, including NW$_b$ and the diffuse emission extending SE, the velocity dispersion is typically relatively low ($\sim 50$\,km s$^{-1}$).

\subsection{GMOS/IFU data analysis: emission line ratios and ionization}\label{sec:ifu_lineratio}

\begin{figure*}
 \centering{}
 \begin{tabular}{ccc}
\multicolumn{2}{c}{\includegraphics[scale=0.4]{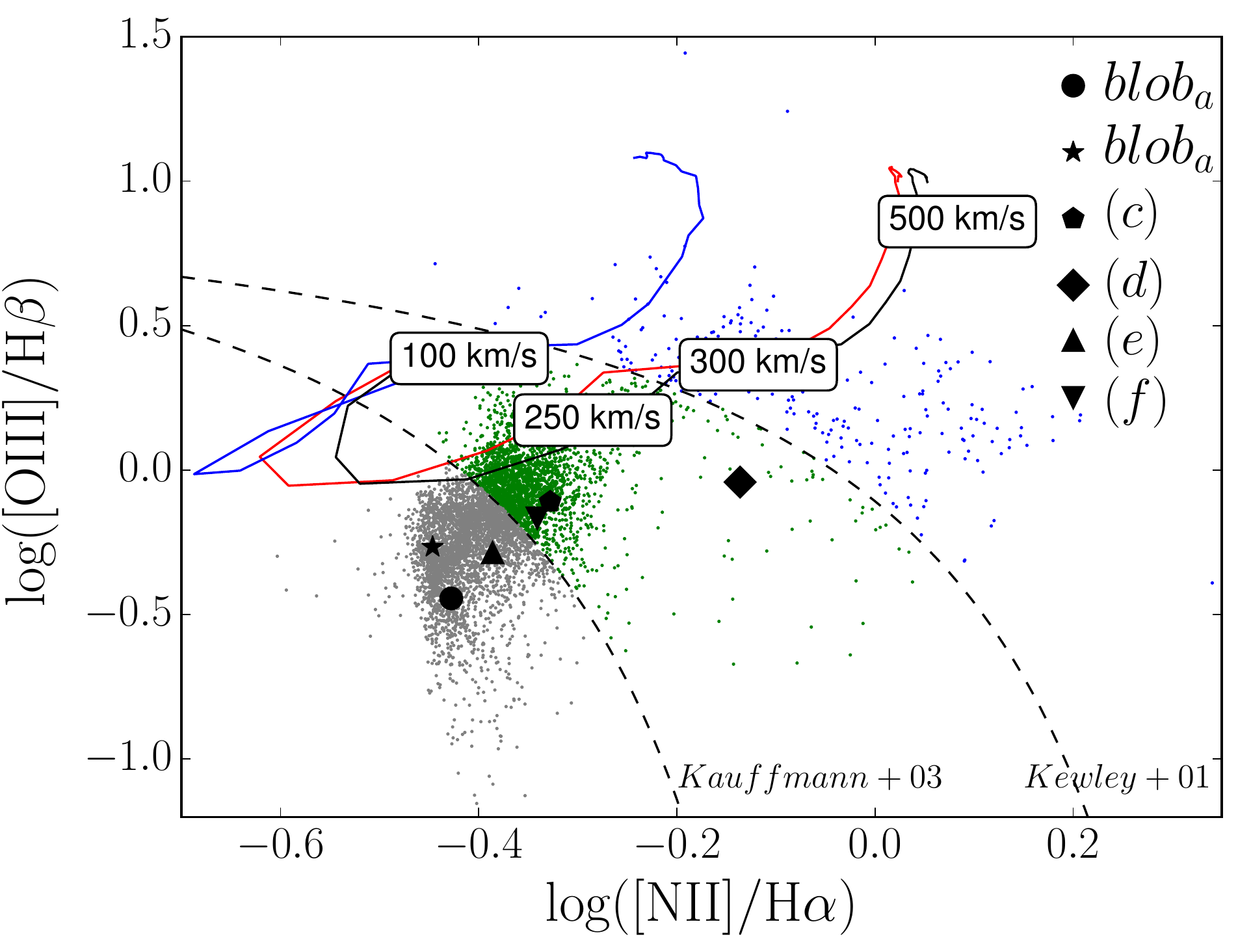}}&
\includegraphics[scale=0.4]{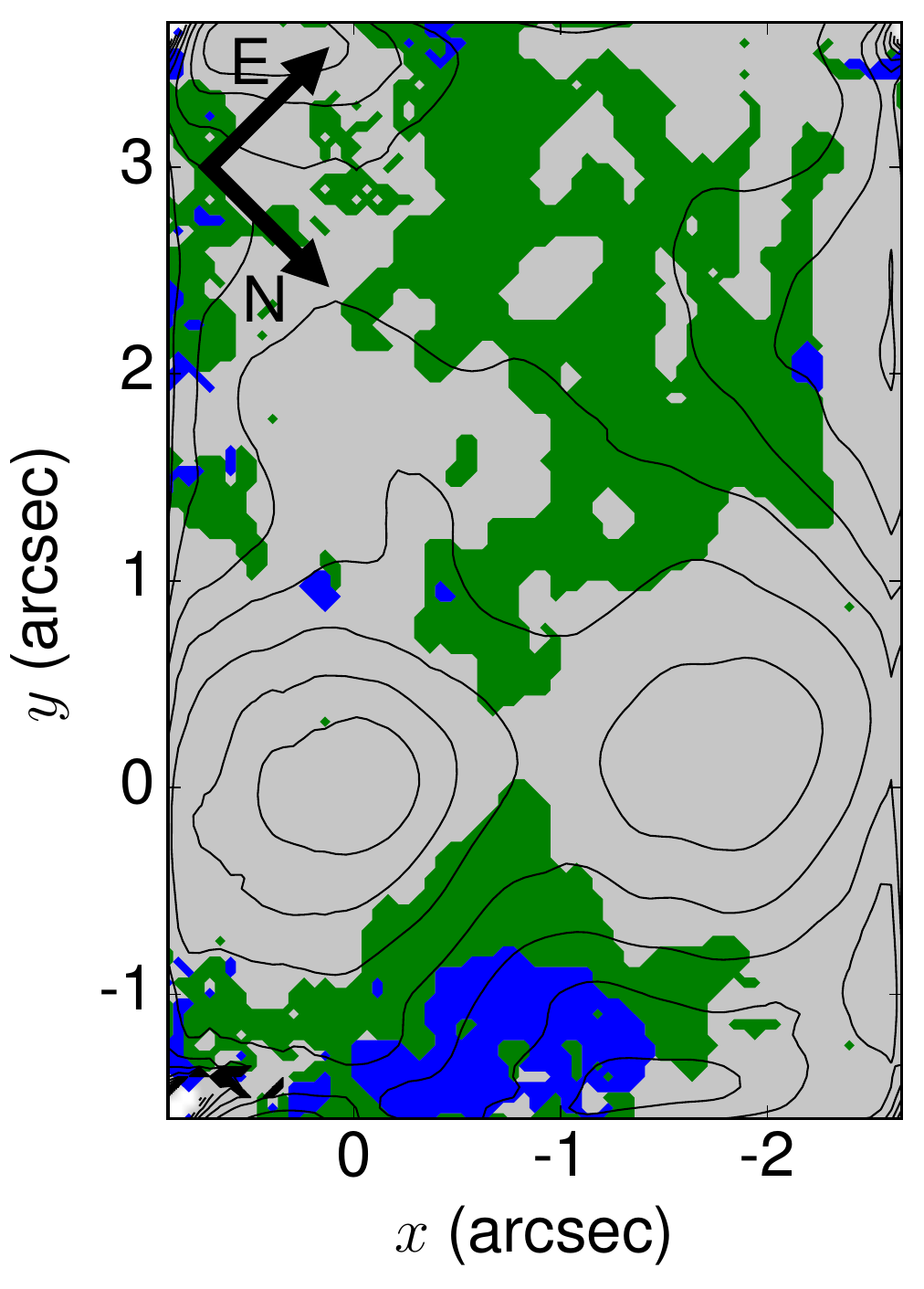}\\
\multicolumn{2}{c}{\includegraphics[scale=0.4]{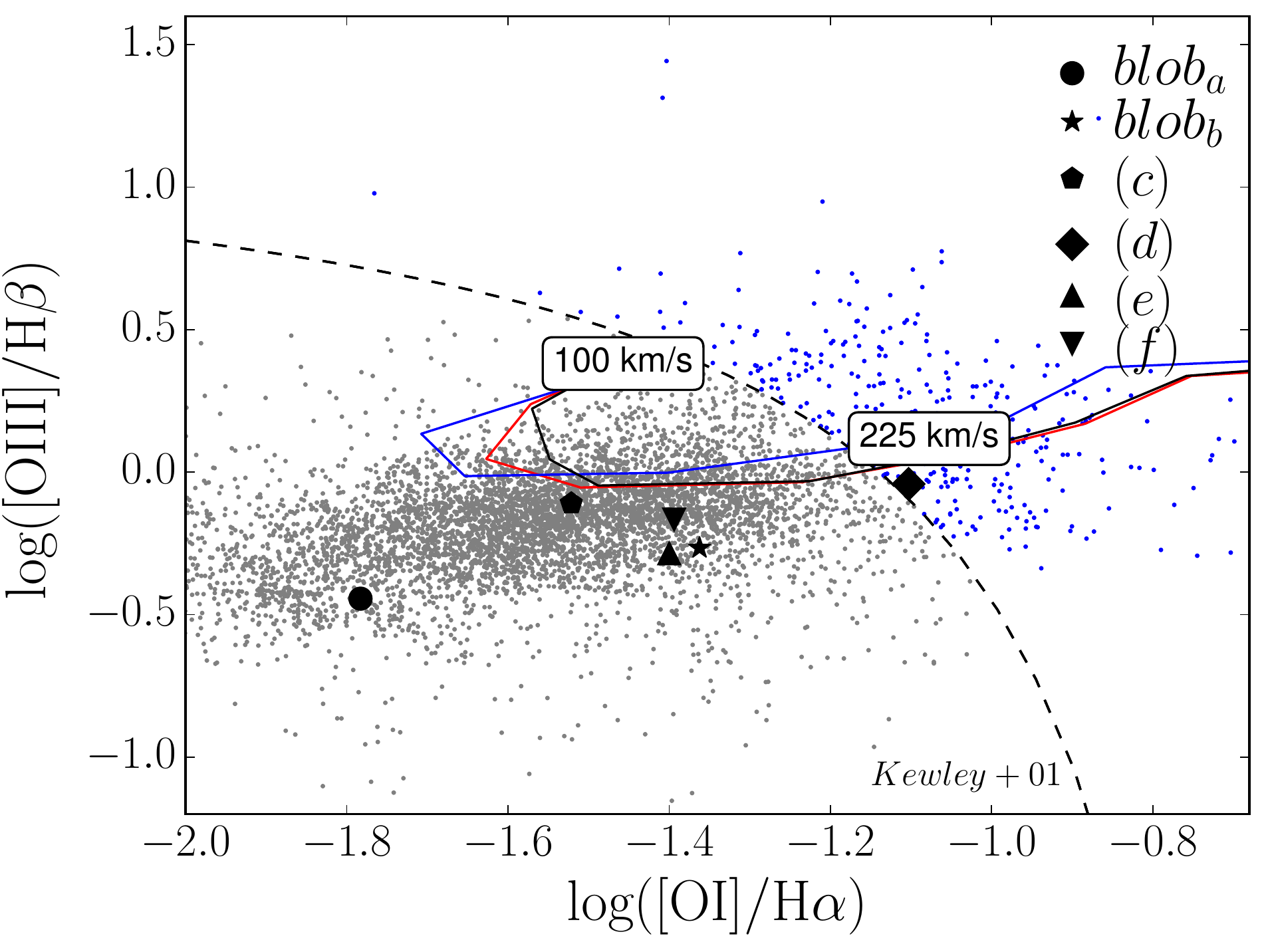}}&
\includegraphics[scale=0.4]{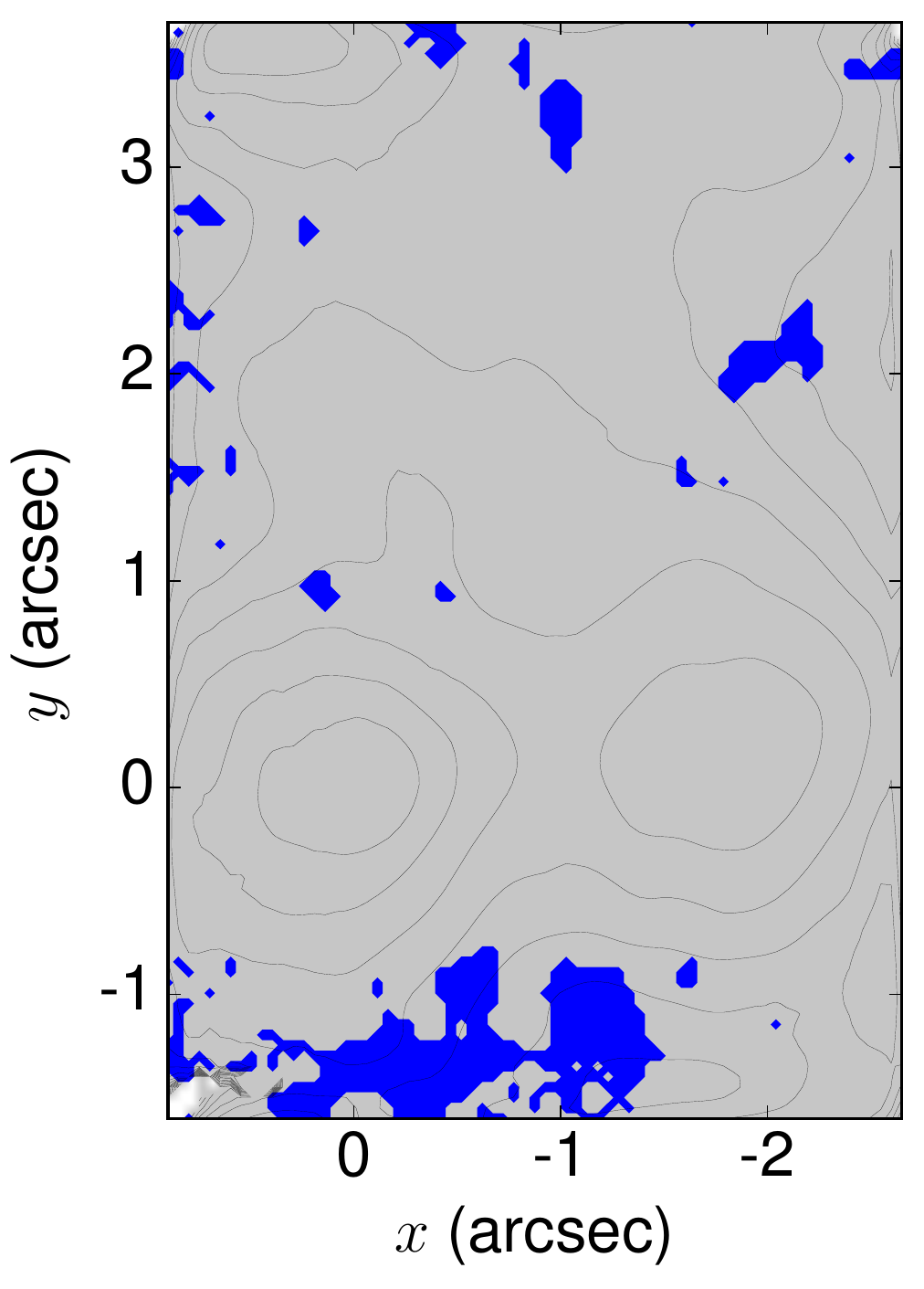}\\
\multicolumn{2}{c}{\includegraphics[scale=0.4]{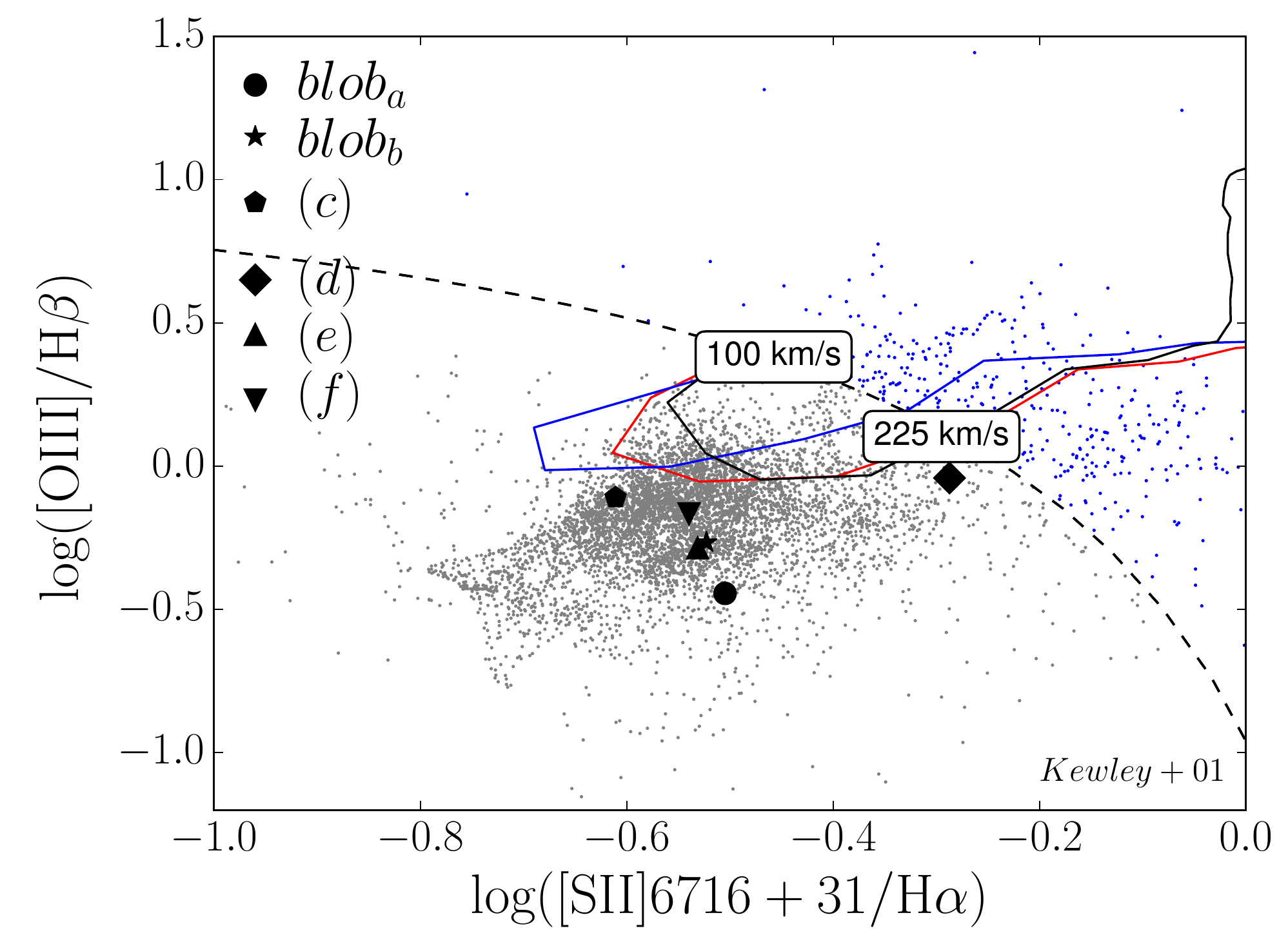}}&
\includegraphics[scale=0.4]{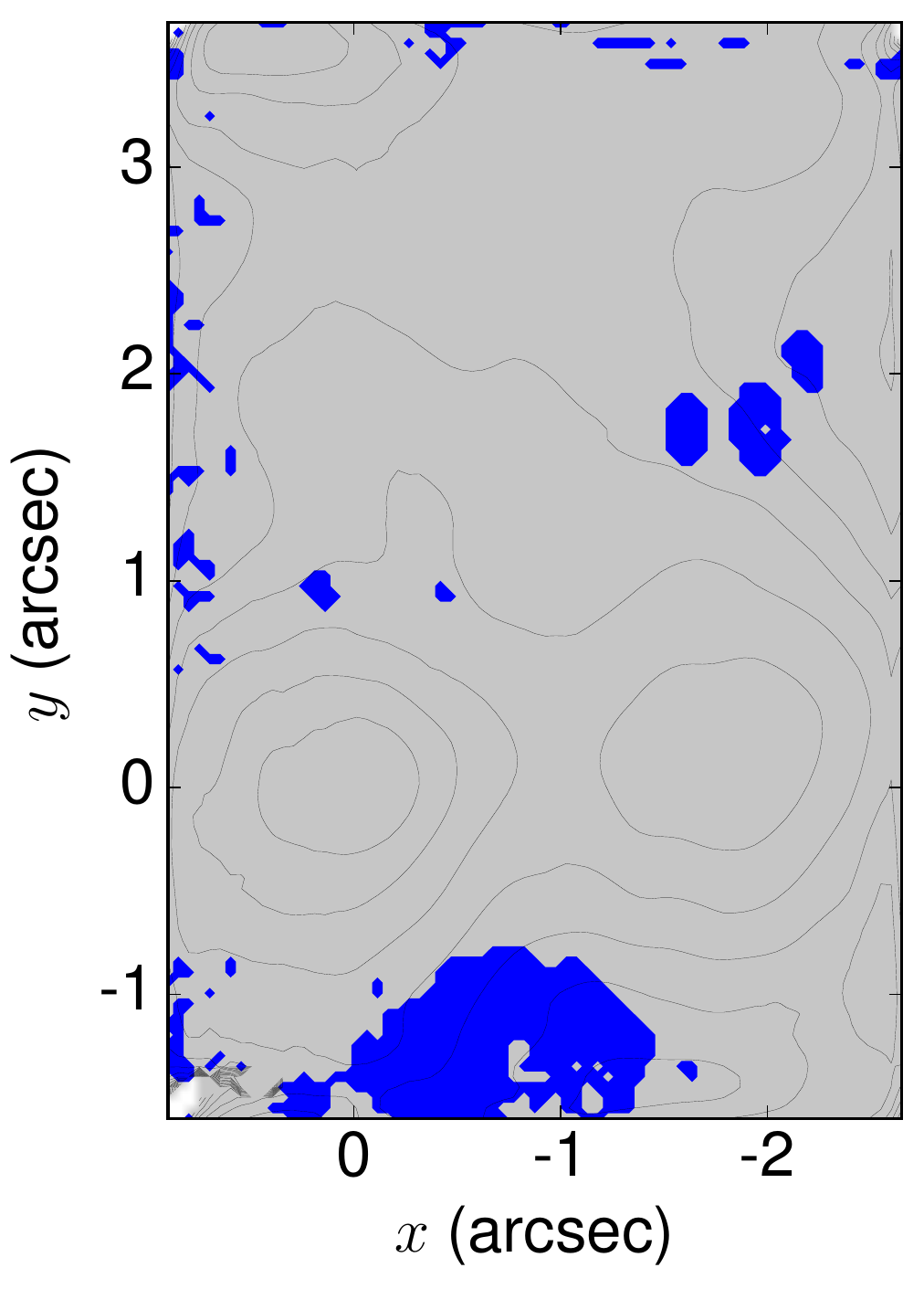}\\
 \end{tabular}
\caption{Left-panels: BPT diagnostic diagrams of spaxel-by-spaxel distribution over 
the whole GMOS/IFU FOV of the NW$_a$. The location of the dividing lines are taken from 
Kewley et al 2001. Green: composites. Gray: H{\sc ii} regions and blue are Seyferts.
Red-solid lines represent shock+precursor models with  solar metallicity, shock 
velocities in the range of 100 to 1000 km s$^{-1}$\, in steps of 25 km s$^{-1}$, magnetic 
parameters B/$\sqrt{n}$ of 0.5$\mu$G, and pre-shock density of 1.0 cm$^{-3}$. Blue and
black solid lines are the same to magnetic parameters B/$\sqrt{n}$ of 10$^{-4}$ $\mu$G and
1.0 respectively. Right-panels: Blue, green and gray points are spatially represented 
by filled regions overplotted on H$\alpha$ map (right-panel).}
\label{fig:bpt-bloba}
\end{figure*}

In order to investigate the ionization mechanisms responsible for the line emission in the region sampled 
by our IFU observations, we have used the line fluxes derived from the Gaussian fits to construct emission 
line ratio diagnostic BPT diagrams \citep{Baldwin1981}. In fact, we use the diagrams introduced by 
\citet{Veilleux1987}: [O\,{\sc iii}]$\lambda\lambda$5007/H$\beta$ versus [N\,{\sc ii}]$\lambda\lambda$6584/H$\alpha$, 
[O\,{\sc iii}]$\lambda\lambda$5007/H$\beta$ versus [O\,{\sc i}]$\lambda\lambda$6300/H$\alpha$, and 
[O\,{\sc iii}]$\lambda\lambda$5007/H$\beta$ versus [S\,{\sc ii}]$\lambda\lambda$6716+31/H$\alpha$, as shown in 
figure \ref{fig:bpt-bloba} (left panels). 
The point cloud in each diagram represents line ratios computed for individual spaxels, using the fluxes of the 
blueshifted or redshifted Gaussian components as appropriate. The symbols represent line ratios derived from spectra 
extracted from apertures centered on regions (a)--(d) as indicated in figure~\ref{fig:espectros}. The dashed lines 
represent the empirical boundaries separating galaxies which have starburst/H\,{\sc ii} region nuclei (grey points), 
active galactic nuclei (blue points) and composite nuclei (green points) derived from analyses of SDSS spectra by
\citet{Kewley2001} and \citet{Kauffmann2003}. The spaxels occupying each region are color-coded and their distributions 
over the IFU FOV are shown in the corresponding excitation maps (right-panels of figure~\ref{fig:bpt-bloba}).  

Also plotted in the BPT diagrams are  loci of radiative shock models taken from the library computed with the 
MAPPINGS III code provided by \citet{Allen2008}. These models include a photoionized precursor and assume solar 
abundances and a pre-shock density 1.0\,cm$^{-3}$. Sequences in shock velocity ranging from 100 to 1000\,km s$^{-1}$ 
are plotted for three values of the magnetic field strength parameter: $B/\sqrt{n} = 10^{-4}, 0.5$ and $1.0\mu$G, 
and a pre-shock density of 1.0 cm$^{-3}$.

\begin{figure}
 \centering
   \includegraphics[scale=0.9,trim=60 360 0 0]{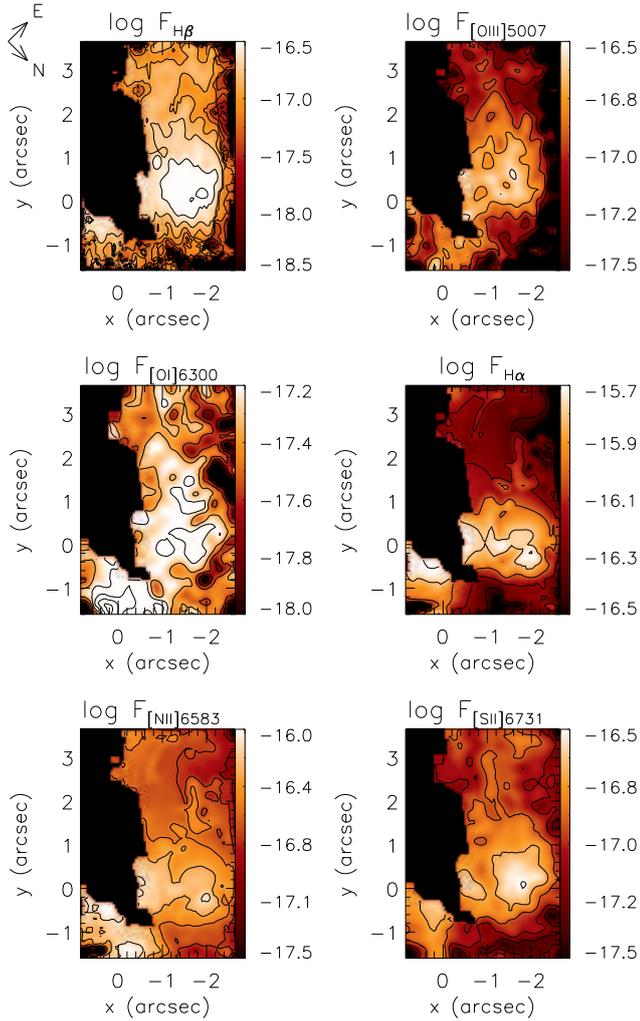}\\
\caption{Total flux maps of the blob$_b$. Flux unit are shown in logarithmic scale of 
erg\,s$^{-1}$\,cm$^{2}$\,fiber$^{-1}$. Black regions demonstrate masked location corresponding 
to the blob$_a$ and/or lower signal-to-noise ratio ($S/N < 3$).}
\label{fig:flux-blobb}
\end{figure}

It is apparent that the vast majority of the spaxels fall within the stellar photoionized ``H\,{\sc ii} region'' zones 
in the [O\,{\sc iii}]$\lambda\lambda$5007/H$\beta$ -- [O\,{\sc i}]$\lambda\lambda$6300/H$\alpha$ and 
[O\,{\sc iii}]$\lambda\lambda$5007/H$\beta$ -- [S\,{\sc ii}]$\lambda\lambda$6716+31/H$\alpha$ diagrams, or in either the 
``H\,{\sc ii} region'' or ``composite''  zones in the [O\,{\sc iii}]$\lambda\lambda$5007/H$\beta$ -- [N\,{\sc ii}]$\lambda\lambda$6584/H$\alpha$. However, the distribution also extends into the ``AGN'' zone in all three diagrams. Notably, the majority of 
the spaxels that fall within the latter zone are located in the region close to the northwestern edge of the FOV that is characterized 
by high velocity dispersion and relatively strong [O\,{\sc i}]$\lambda$6300 emission (e.g., figures~\ref{fig:flux-blobb} 
and \ref{fig:vel-blobb}). 

The line ratios derived from the integrated spectra of regions (a)--(c) and (d)--(f) generally cluster quite closely together within the  ``H\,{\sc ii} region'' zones (or close to the ``H\,{\sc ii} region''--``Composite'' boundary in the O\,{\sc iii}]$\lambda\lambda$5007/H$\beta$ -- [N\,{\sc ii}]$\lambda\lambda$6584/H$\alpha$ diagram). The exception is region (d), which falls close to
the ``H\,{\sc ii} region''--``AGN'' or ``composite''--``AGN'' boundaries in all three diagrams. This region is also located near the northwestern edge of the FOV and has a relatively high velocity dispersion.

The distribution of spaxels and selected regions in the BPT diagrams indicates that stellar photoionization is the dominant ionization mechanism over most of the FOV, including the bright blobs, NW$_a$ and NW$_b$, and the more diffuse emission covering most of the rest of the field. However, the region northwestern of the bright blobs and centered roughly between them along the $x$-axis is characterized by a high velocity dispersion and comparatively large values of the [N\,{\sc ii}]$\lambda$6584/H$\alpha$, 
[O\,{\sc i}]$\lambda$6300/H$\alpha$ and [S\,{\sc ii}]$\lambda$6716+31/H$\alpha$ ratios. The line ratios in this region are consistent with weak AGN photoionization (i.e., LINER-like) but given the high velocity dispersion, a more likely explanation is shock ionization. The shock models indeed roughly reproduce the lateral scatter towards the ``AGN'' zone in the BPT diagrams and in the O\,{\sc iii}]$\lambda\lambda$5007/H$\beta$ -- [O\,{\sc i}]$\lambda\lambda$6300/H$\alpha$, and [O\,{\sc iii}]$\lambda\lambda$5007/H$\beta$ 
-- [S\,{\sc ii}]$\lambda\lambda$6716+31/H$\alpha$ diagrams, at least, the region (d) line ratios are consistent with a shock velocity of $\sim 200$ km\,s$^{-1}$. On the other hand, the models do not match the observed line ratios particularly well in the [O\,{\sc iii}]$\lambda\lambda$5007/H$\beta$ -- [N\,{\sc ii}]$\lambda\lambda$6584/H$\alpha$ diagram, because the models under predict the [N\,{\sc ii}]$\lambda\lambda$6584/H$\alpha$ ratio for a given shock velocity. 

Nevertheless, it seems reasonable to conclude that the line emission in region (d) and in the blue-shaded regions northwestern of the bright blobs in figure~\ref{fig:bpt-bloba} predominantly arises due to shock ionization. As can be seen in the excitation map corresponding to the [O\,{\sc iii}]$\lambda\lambda$5007/H$\beta$ -- [N\,{\sc ii}]$\lambda\lambda$6584/H$\alpha$ diagram, the ``shock-ionized'' region is surrounded by green-shaded ``composite'' region spaxels. This may indicate a transition zone where both shocks and H\,{\sc ii} regions contribute to the spectrum.

\begin{table}
\centering
\renewcommand{\tabcolsep}{1mm}
\caption{Line Ratios derived from GMOS/IFU data.}\label{lineratio}
\begin{tabular}{lcccc}
\hline\hline
\noalign{\smallskip}
\noalign{\smallskip}
&  log($\frac{H\alpha}{[N\,II]}$) & log($\frac{[O\,III]}{H\beta}$) & log($\frac{[O\,I]}{H\alpha}$) & log($\frac{[S\,II]}{H\alpha}$)\\
\hline
Blob$_a$ & -0.42 & -0.44 & -1.78 & -0.50 \\
Blob$_b$ & -0.45 & -0.26 & -1.36 & -0.52 \\
Region (c) & -0.33 & -0.11 & -1.52 & -0.61 \\
Region (d) & -0.14 & -0.04 & -1.10 & -0.29 \\
Region (e) & -0.39 & -0.28 & -1.40 & -0.53 \\
Region (f) & -0.34 & -0.17 & -1.39 & -0.54 \\
\noalign{\smallskip}
\hline
\end{tabular}
\end{table}

\subsection{H$\alpha$ photometry and star formation rates}\label{result-HaPhot}

As star formation appears to dominate the line emission from most of the IFU FOV, including the two brightest blobs 
(Section~\ref{sec:ifu_lineratio}), H$\alpha$ fluxes measured from the continuum subtracted FR656N ramp filter image 
were used to estimate star formation rates. Flux measurements were made for the entire IRAS17526+3253 system and 
also within smaller apertures centered on each of the two merging galaxies, the IFU field and blobs NW$_a$ and NW$_b$ within 
the IFU field. The aperture sizes used for the photometric measurements are listed in Table~\ref{reddening}. 
In order to determine the intrinsic H$\alpha$ fluxes from these measurements it is necessary to correct for both dust 
extinction and the contribution of the  [N\,{\sc ii}]$ \lambda\lambda$6548,6584 lines, which fall within the filter passband. 
Approximate corrections were derived using line intensity ratios obtained from the IFU data. As there are large changes 
in the Balmer decrement over the IFU field of view, we derived two values for the H$\alpha$ extinction, one using the 
H$\alpha$ and H$\beta$ line fluxes integrated over the whole FOV, and the other using average of the line fluxes measured 
in 1$\arcsec$ apertures centered on blobs NW$_a$ and NW$_b$. For the IFU/FOV, we find $I_{H\alpha}/I_{H\beta}\approx 8.0$, 
yielding $A_{H\alpha}\approx 2.7$, using the standard Milky Way extinction curve (\citealp*{Cardelli1989}; $R_V = 3.1$) 
and assuming the Case B recombination value, $I_{H\alpha}/I_{H\beta}= 2.87$, for the intrinsic Balmer decrement 
\citep{Osterbrock2006}. For the blobs, the average Balmer decrement is $I_{H\alpha}/I_{H\beta}\approx 4.7$, implying 
a much lower extinction, $A_{H\alpha}\approx$ 1.3.

\begin{table}
\centering
\renewcommand{\tabcolsep}{0.9mm}
\caption{H$\alpha$ luminosities and SFRs measured from HST images.}\label{reddening}
\begin{tabular}{lcccc}
\hline\hline
\noalign{\smallskip}
\noalign{\smallskip}
Flux	&	Northwestern	&	East	&	System 	&	IFU/FOV	\\
Aperture radius & 5\farcs & 5\farcs & 27\farcs & 5\farcs1 x 3\farcs4 \\
\hline
\multicolumn{5}{c}{Reddening and [N\,{\sc ii}] correction derived from IFU/FOV}\\
\hline
F$_{\rm H\alpha+[N\,{\sc II}]}$ (erg/s/cm$^{2}$)	&	1.4e-13	&	8.3e-14	&	6.0e-13	&	1.7e-14	\\
F$_{\rm H\alpha, corr}$	&	1.1e-12	&	6.1e-13	&	4.5e-12	&	4.0e-13	\\
L$_{\rm H\alpha}$ (erg/s)	&	1.5e+42	&	8.5e+41	&	6.2e+42	&	5.6e+41	\\
SFR (M$_{\odot}$/yr)	&	7.9	&	4.5	&	33.1	&	2.98	\\
\hline
\multicolumn{5}{c}{Reddening and [N\,{\sc ii}] correction derived using}\\
\multicolumn{5}{c}{an average value from IFU NW$_a$ and NW$_b$}\\
\hline
F$_{\rm H\alpha+[N\,{\sc II}]}$ (erg/s/cm$^2$)	&	1.4e-13	&	8.3e-14	&	6.0e-13	&	1.7e-14	\\
F$_{\rm H\alpha, corr}$	&	3.5e-13	&	2.0e-13	&	1.5e-12	&	1.3e-13	\\
L$_{\rm H\alpha}$ (erg/s)	&	4.9e+41	&	2.8e+41	&	2.0e+42	&	1.8e+41	\\
SFR (M$_{\odot}$/yr)	&	2.58	&	1.48	&	10.8	&	0.97	\\
\hline\hline
\noalign{\smallskip}
\end{tabular}
\end{table}

Inspection of the FR656N image suggests that it is reasonable to suppose that the extinction affecting the IFU field is fairly representative of the emission line regions elsewhere in the system. Therefore, assuming that the two values of A$_{H\alpha}$ derived from the IFU data represent upper and lower limits on the amount of extinction affecting a given region, we have
used both values to calculate extinction corrections for each component, with the exception of the blobs themselves for which only the corresponding local value of A$_{H\alpha}$ was used.

After applying extinction corrections, values of the [N\,{\sc ii}]$\lambda$6584/H$\alpha$ ratio measured from the IFU field were used to estimate and remove the contribution of the
 [N\,{\sc ii}]$ \lambda\lambda$6548,6584 lines to the fluxes measured from the HST images. Again, separate values of this ratio were derived, corresponding to the
the whole IFU/FOV, and the average of the two blobs. Assuming $I_{[N\,{\sc II}]\lambda 6584}/I_{[N\,{\sc II}]\lambda 6548} = 2.9$, we estimate H$\alpha$ contributions to the extinction corrected fluxes of 65\% and 73\%, respectively. The measured  $H\alpha+$[N{\sc ii}] fluxes, the reddening corrected H$\alpha$ fluxes and the H$\alpha$ luminosities are listed for each region for both the ``high'' and ``low'' extinction cases in Table~\ref{reddening}.

The corresponding star formation rates are also listed in Table~\ref{reddening} and were estimated from the H$\alpha$ luminosity using the relationship given by \citet{Calzetti2007}:
SFR$_{H\alpha}(M_{\odot}\,yr^{-1})\,=\,5.3\,\times\,10^{-42}\,L_{H\alpha}$. The measured H$\alpha$ luminosities imply that the {\em unobscured} star formation rate for the entire IRAS17526+3253  
system is $\sim 10 - 30$\,M$_{\odot}$yr$^{-1}$, with the central regions of the interacting galaxies contributing $\sim 2.6 - 7.9$\,M$_{\odot}$yr$^{-1}$  and $\sim 1.5 - 4.5$\,M$_{\odot}$yr$^{-1}$ for the northwestern and east components, respectively, and with most of the star formation in the latter being associated with the bright H\,{\sc ii} regions within the IFU/FOV.

\section{Discussion}\label{sec:discussion}

\subsection{Large-scale morphology and merger stage}\label{sec:dis-morph}

Our HST/ACS observations confirm earlier suggestions \citep{Andreasian1994,Wagner2013} that IRAS17526+3253 is a late stage major merger. 
Although the northwestern nucleus appears larger and brighter in the F814W image and has a brighter radio source, the K$_{s}$ magnitudes derived from the 2MASS image suggest that 
the two galaxies have similar bulges, with a mass ratio $\sim 1$. Both galaxies appear to be quite highly inclined and the elongated nature of the 
system as a whole suggests that it is viewed close to the orbital plane of the interaction. The two galaxy nuclei are separated by a projected distance of $\sim 9$\,kpc, and appear to be embedded within   
a complex, disturbed structure with multiple dust lanes, large-scale tidal tails or plumes and widely distributed clumps of star formation. 
A comparison with merger simulations \citep[e.g.][]{Lotz2008,Hung2016} suggests that the morphological properties of the system are consistent with an mid to advanced merger stage, probably between the second pericenter passage and final coalescence.

Both nuclei host compact, but resolved radio sources at 1.49\,GHz; which have properties consistent 
with radio emission from starburst galaxies \citep{Baan2006}. 
An optical spectrum obtained by \citep{Baan1998}, is also consistent with starburst activity, but this classification probably relates 
only to the NW nucleus (private communication).

Our IFU observation was designed to sample the brightest emission line region in the northwestern galaxy. As discussed in \ref{sec:results}, this region has a complex 
structure with at least two kinematically distinct components, but over most of the field of view, the emission line ratios are consistent with photoionization in star formation regions.
However, the radio source associated with this galaxy is centered within the prominent dust lane that borders the northern corner of the IFU field. Thus, assuming 
that the radio source core corresponds to the nucleus of the galaxy, it is evidently embedded in the dust lane and heavily obscured at optical wavelengths.
It is likely, therefore, that the IFU field does not include the nucleus itself, but samples the surrounding star formation regions. 

The Imperial IRAS-FSC Redshift Catalogue \citep[IIFSCz][]{Wang2009} gives a far infrared luminosity of $L_{FIR} = 10^{11.19} $\,L$_{\odot}$ for the IRAS17526+3253 system, implying
a global star formation rate $SFR_{FIR} \sim 27$M$_{\odot}$\,yr$^{-1}$. This is approximately consistent with the total of the obscured and unobscured star formation rates estimated from the 1.4\,GHz radio sources \citep{Baan2006}
and the continuum-subtracted H$\alpha+[NII]$ image, respectively (Section~\ref{result-HaPhot}). 
The FIR luminosity and global star formation rate are both somewhat higher than the median values ($\sim 10^{11}L_{\odot}$ and $\sim 15$\,M$_{\odot}$\,yr$^-1$, respectively) 
for the sample of $\sim 3000$ local mergers studied by \citep{Carpineti2015}, but are very close to the peaks of the respective distributions. 
Evidently, IRAS17526+3253 is a fairly typical example of a spiral-spiral merger in the local universe, insofar as its star formation rate is concerned.

\begin{table}
\centering
\renewcommand{\tabcolsep}{1mm}
\caption{Star Formation Rate of IRAS17526+3253.}\label{sfr_all}
\begin{tabular}{lcccccc}
\hline\hline
\noalign{\smallskip}
\noalign{\smallskip}
Flux	& Northwestern	& East	& System & NW$_a$ & NW$_b$ & IFU/FOV \\
\hline
H$_\alpha$ (HST)	& 7.9	& 4.5	& 33.1	& 0.3	& 0.2	& 3.0 \\
H$_\alpha$ (IFU)	& - & - & - & 4.1	& 2.8	& 12.0 \\
L$_{FIR}$$^a$ & - & - & 26.8	& - & - & - \\
L$_{1.4GHz}$$^b$ & 13.1	& 6.0 & - & - & - & - \\
L$_{1.4GHz}$$^a$ & - & - & 13.7	& - & - & - \\
\hline
\multicolumn{7}{l}{$^a$IIFSCz, \citet{Wang2009}}\\
\multicolumn{7}{l}{$^b$ \citet{Baan2006}}\\
\noalign{\smallskip}
\end{tabular}
\end{table}

\subsection{Ionized gas excitation and kinematics}\label{sec:dis-morph}

As noted above, the IFU field samples a bright emission line region within the northern galaxy, bordering 
the compact radio source that we presume locates the obscured nucleus. As discussed in Section~\ref{sec:ifu_results}, 
the emission line region is divided into two main components by a discontinuity in the velocity field of 
magnitude $\sim 200$\,km\,s$^{-1}$, one of which (on the SW side) is blue-shifted and the other (NE side) 
red-shifted, relative to the mean velocity. These components are also present in the velocity channel maps and 
are also clearly visible in the PCA tomograms as eigenvectors 1 and 2, which together account for 95\% 
of the variance. In addition, the increased velocity dispersion and line splitting along the boundary 
defined by the velocity field discontinuity are also consistent with physically distinct 
components at different velocity shifts, with one slightly overlapping the other in projection. 
Morphologically, each kinematic component is dominated by a bright blob (NW$_a$ and NW$_b$, respectively), 
appearing both in continuum and in line emission, which  are separated by $\sim$850\,pc, and surrounded by more diffuse ionized gas. 

The emission line ratios indicate that star formation is the main source of the line emission in both kinematic components, including the two bright blobs.
However, the fainter line emission bordering the blobs to the northwest is characterized by a higher velocity dispersion and line ratios consistent  
with shock ionization for shock velocities $\sim 200$ km\,s$^{-1}$. This shock ionized region is situated between the bright blobs and both
the nuclear radio source and the prominent dust lane, which crosses the northern corner of the IFU field. 
 
The origin of the two kinematic components is unclear. However, it seems plausible that one is associated 
with the disrupted disk of the northwestern galaxy, while the other is tidal debris, seen in projection and partially overlapping the disk. It is possible that this ``debris'' is a tidal
tail originating in the southeastern galaxy since the longslit spectrum obtained by \citet{Andreasian1994} suggests that this galaxy is redshifted by $\sim 300$\,km\,s$^{-1}$ relative to its NW counterpart. 
The HST F814W image suggests that the prominent dust lane crossing the core of NW galaxy may also be associated with the tidal tail.
In this scenario, the shocked gas may be the result of an interaction between the tidal tail and part of the NW galaxy's disk.

\subsection{Maser sources}\label{sec:dis-maser}

As noted in Section~\ref{sec:introd}, the identification of an OH maser source in IRAS17526+3253 rests on the feature reported by \citet{Martin1989b}, which has a velocity 
 $\approx 7500$km\,s$^{-1}$. In comparison, the H\,I 21\, cm emission observed by the same author peaks  
 at $\approx 7800$km\,s$^{-1}$, approximately the same velocity as the H$_{2}$O maser lines \citep{Wagner2013}. In addition, CO emission has been detected
both at $\approx 7500$km\,s$^{-1}$ and $\approx 7800$km\,s$^{-1}$ \citep{Baan2008}. There are evidently two velocity systems present in the IRAS17526+3253 system,
which are most probably associated with the two galaxies revealed by the HST image. The OH maser emission and the blueshifted component of the CO(2-1) line have velocities 
comparable with that of the blueshifted component of ionized gas, which we have identified with the NW galaxy. As discussed above (Section~\ref{sec:dis-morph}), the SE galaxy 
appears to be redshfted by $\sim 300$\,km\,s$^{-1}$ with respect to the latter. Based on the kinematics, therefore, we can conclude that the OH maser source is located in the
NW galaxy, whereas the H$_{2}$O masers are associated with the SE galaxy.

\subsection{Comparison with previously studied OHMG}\label{sec:comp_ohmg}

IRAS17526+3253 is the fourth OH megamaser galaxy that we have studied in detail to date. These systems exhibit a range of morphological and gas 
excitation characteristics. Multiwavelength imaging and spectral energy distribution analysis of IRAS16399-0937 \citep{Sales2015} revealed that it is an mid to advanced merger with two nuclei
embedded in a tidally distorted envelope, with a total SFR $\sim 20$\,M$_{\odot}$\,yr$^{-1}$. 
The nuclei are separated by $\sim 3.4$\,kpc and the northwestern hosts a dust embedded AGN of luminosity $L_{\textnormal{bol}}\sim 10^{44}$\,erg\,s$^{-1}$.
IRASF23199+0123 was studied using HST and VLA images as well Gemini IFU spectroscopy \citep{Heka2018a}. This is an interacting pair, connected by a tidal tail, 
whose two galaxies are separated by 24\,kpc. The detection of a broad H$\alpha$ emission line revealed that the eastern member of the pair hosts a Seyfert 1 nucleus.
In this, case we were also able to obtain VLA maps the OH maser emission, showing that the eastern galaxy also hosts two masing sources, which appear to be closely associated with the AGN. 
The OH masers are located in the vicinity of a region of enhanced velocity dispersion and higher [NII]/H$\alpha$ ratios, 
suggesting that they are associated with shocks driven by AGN outflows. In contrast, IRAS03056+2034 is a barred spiral with morphological irregularities suggesting interactions 
and a circum-nuclear ring (radius $\sim 0.5$\,kpc) of star-forming regions, with an SFR $\sim 5$\,M$_{\odot}$\,yr$^{-1}$ 
\citep{Heka2018b}. As in IRAS17526+3253, there is no clear evidence that the galaxy hosts an AGN.

These four systems are too small a sample upon which to base any firm conclusions. However, it is perhaps notable that OH masers are present in 
mid to advanced major mergers (IRAS16399-0937, IRAS17526+3253) as well as in a galaxy that may be in an early stage of a possible merger (IRASF23199+0123) and one that
is not undergoing any noticeable interaction with a similarly sized companion (IRAS03056+2034). Two of these systems harbour moderately luminous embedded AGN in at least one of their components 
(IRAS16399-0937 and IRASF23199+0123), 
but the other two appear to be dominated by star formation. In the one case for which we have so far been able to establish the location of the OH maser sources, they are associated with the AGN, perhaps
arising in shocks driven by the latter.

IRAS17526+3253 is one of only a handful of galaxies known to host both luminous OH and H$_{2}$O masers  \citet{Wiggins2016}. Of these, only two are dual 
megamaser hosts, Arp\,299, \citep{Tarchi2011} and II\,Zw\,96, \citep{Wagner2013, Wiggins2016} (a possible third candidate, UGC5101, has a dubious OH detection; \citeauthor{Wiggins2016}).
The apparent dearth of systems hosting megamasers of both species may be due to the different physical conditions required for maser emission in each case, with OH masers typically occurring in (U)LIRG systems, 
whereas H$_{2}$O masers are usually associated with AGN \citep{Lo2005}. The two dual megamaser hosts are both merging systems with spatially distinct nuclei and 
both are spectroscopically classified as H\,II galaxies. Although IRAS\,7526+3253 does not quite qualify as an OH megamaser host by the criteria adopted by \citeauthor{Wiggins2016} ($L_{OH}> 10L_\odot$), it's OH maser luminosity is only slightly less than that of Arp\,299. On the other hand, IRAS\,7526+3253 has an H$_{2}$O maser luminosity comparable with that of the other two systems. In several respects, therefore, IRAS\,17526+3253 is similar to the two known dual megamaser hosts. In Arp\,299, the H$_{2}$O maser emission was detected in the nucleus that also hosts the OH maser source \citep{Tarchi2011}. The II\,Zw\,96 system appears to be a complex advanced merger in which the OH maser source is associated with a tidally stripped and heavily reddened nucleus that may contain an embedded AGN \citep{Migenes2011}, but it is not clear if the H$_{2}$O maser comes from the same region. In IRAS\,7526+3253, the evidence suggests that the OH and H$_{2}$O masers are associated with different nuclei (NW and SE, respectively). There is no evidence supporting the presence an AGN in either nucleus, although this possibility cannot be ruled based on the existing data, and we currently lack a spectroscopic classification for the SE nucleus. Thus, further observations of IRAS\,17526+3253, in particular, to determine if either nucleus hosts an AGN may shed light on any connections between OH and H$_{2}$O masers and galaxy merger stage.


\section{Summary and Conclusion}\label{sec:summary}

We have presented a multiwavelength study of the luminous infrared galaxy IRAS17526+3253, using 
Gemini GMOS/IFU, HST, 2MASS and VLA data. 
Previous radio studies have classified the two major components of the IRAS17526+3253
system as starburst galaxies. This system has also been reported to host OH megamaser emission, although 
strong radio frequency interference frustrated a more recent attempt to confirm the detection.

The data set analyzed here combines new emission line  (H$\alpha$+[N{\sc ii}) and broad-band (FR914M, F814W)
imaging obtained with the HST ACS Wide Field Camera with archival VLA radio imaging. We also present  
a two-dimensional analysis of the gas excitation and kinematics of a bright emission line region in the
northwestern galaxy. The main results from this analysis are summarized below:

\begin{itemize}

\item The HST/ACS, 2MASS Ks-band and 1.49~GHz VLA images clearly reveal a mid-stage major merger with 
two main galaxy nuclei separated by a projected distance of $\sim$8.5~kpc, embedded in an elongated irregular 
envelope. The northwestern galaxy is quite highly inclined, with the second component 
appearing to be merging with its companion and probably between the second pericenter passage and final 
coalescence. The morphological elongation of the system suggests that we are
observing it from a viewpoint quite  close to the orbital plane. The two nuclei are clearly distinguished in the 
near-IR (2MASS Ks-band) and each is associated with a compact (but resolved) 1.49\,GHz radio  source, previously 
attributed to star formation.  
The HST/ACS F814W image shows that the two galaxies are embedded in a large tidally distorted envelope 
with a complex structure including numerous dust lanes and bright knots.

\item The HST H$\alpha$+[N{\sc ii} emission line image 
shows at least 27 compact regions of ongoing star formation across the envelope over scales of several 
10's of kpc.
IRAS17526+3253 has previously been classified, based on an optical spectrum as a star-forming galaxy. 

\item Our Gemini/IFU observation was designed to sample the brightest emission line region in the northern galaxy, which lies just outside the galaxy nucleus.
Assuming that the radio core corresponds to the position of the latter, it is evidently obscured by a prominent dust lane 
that borders the northern corner of the Gemini/IFU FOV. Two distinct kinematic components, divided by a discontinuity 
in the velocity field of magnitude $\sim$ 200 km s$^{-1}$, are present in the IFU field. One component (on the SW side) is blue-shifted 
and the other (NE side) red-shifted, relative to the mean velocity. These components are also present in the 
velocity channel maps and are also clearly visible in the PCA tomograms. Each 
kinematic component is dominated by a bright blob (NW$_a$ and NW$_b$, respectively),  appearing both
in optical continuum and in line emission, which are separated by $\sim$850 pc, and surrounded by more diffuse
ionized gas.

\item The emission line ratio diagnostic diagrams indicate that star formation is the main source of the line 
emission in both kinematic components, including the two bright blobs. However, the fainter line emission bordering 
the blobs to the northwestern is characterized by a higher velocity dispersion and line ratios consistent with shock ionization 
for shock velocities $\sim$ 200 km s$^{-1}$. This shock ionized region is situated between the bright
blobs and both the nuclear radio source and the prominent dust lane crossing the northern corner of the field.

\item We suggest that the two kinematic components represent, on the SW side of the IFU field, the disrupted disk of the northwestern galaxy, and on the NE side,
tidal debris, seen in projection and partially overlapping the disk. This material may be part a tidal tail from the southeastern galaxy.
The shocked gas may be the result of an interaction between the tidal tail and part of the NW galaxy's disk.

\item The measured H$\alpha$ luminosities imply that the unobscured star formation rate of the whole
IRAS17526+3253 system is $\sim 10 - 30$\,M$_{\odot}$yr$^{-1}$, with the central regions of the interacting galaxies 
contributing $\sim 2.6 - 7.9$\,M$_{\odot}$yr$^{-1}$  and $\sim 1.5 - 4.5$\,M$_{\odot}$yr$^{-1}$ for the northwestern and 
east components, respectively, and with most of the star formation in the northwestern being associated with the bright 
H\,{\sc ii} regions within the IFU/FOV.

\item IRAS17526+3253 is one of only a few systems known to host luminous OH and H$_{2}$O masers. The velocities of the OH and H$_{2}$O masers suggest that they are associated 
with the NW and  SE galaxies, respectively.

\end{itemize}

Support for program HST-SNAP 11604 was provided by NASA through a grant from the Space Telescope Science Institute, which is operated by the Association of Universities for Research in Astronomy, Inc., under NASA contract NAS 5-26555. This  material  is based upon work partly supported by the National Aeronautics and Space Administration under  Grant  No.   NNX11AI03G  issued  through  the  Science  Mission  Directorate. D.  A. Sales gratefully acknowledge for partial financial support received from FAPERGS/CAPES n.05/2013 and CNPq Universal 01/2016. A. R and D. A. Sales acknowledge the Sociedade Brasileira de F\'isica (SBF) and the American Physical Society (APS) for financial support received from the Brazil-U.S. Exchange Program. The work of S.B. and C.O. was supported by NSERC (Natural Science and Engineering Research Council of Canada). Based on observations obtained at the Gemini Observatory [include additional acknowledgement here, see section 1.2], which is operated by the Association of Universities for Research in Astronomy, Inc., under a cooperative agreement with the NSF on behalf of the Gemini partnership: the National Science Foundation (United States), the National Research Council (Canada), CONICYT (Chile), Ministerio de Ciencia, Tecnolog\'ia e Innovaci n Productiva (Argentina), and Minist\'{e}rio da Ci\^{e}ncia, Tecnologia e Inovaca\c{c}\~ao (Brazil). This research has made use of the NASA/IPAC Extragalactic Database (NED) which is operated by the Jet Propulsion Laboratory, California Institute of Technology, under contract with the National Aeronautics and Space Administration.











\bsp	
\label{lastpage}
\end{document}